\documentclass[12pt]{article}
\usepackage{amssymb, amsmath, hyperref, verbatim}
\usepackage{graphicx}
\usepackage[latin1]{inputenc}
\def\ltwid{\mathrel{\raise.3ex\hbox{$<$\kern-.75em\lower1ex\hbox{$\sim$}}}}
\def \be{\begin{equation}}
\def \ee{\end{equation}}
\def \bea{\begin{eqnarray}}
\def \eea{\end{eqnarray}}

\def \f{\frac}

\def\ltwid{\mathrel{\raise.3ex\hbox{$<$\kern-.75em\lower1ex\hbox{$\sim$}}}}

\def\square{\kern1pt\vbox{\hrule height 1.2pt\hbox{\vrule width 1.2pt\hskip 3pt
   \vbox{\vskip 6pt}\hskip 3pt\vrule width 0.6pt}\hrule height 0.6pt}\kern1pt}

\begin{document}

\begin{titlepage}
\begin{flushright}
SPIN-09/18, ITP-UU-09/18
\end{flushright}

\vspace{0.5cm}

\begin{center}
\bf{Regulating the infrared by mode matching:\\
A massless scalar in expanding spaces with constant deceleration}
\end{center}

\vspace{0.3cm}

\begin{center}
T. M. Janssen$^{*}$, T. Prokopec$^{\dagger}$
\end{center}
\begin{center}
\it{Inst. for Theor. Physics \& Spinoza Institute,
Utrecht University,\\
Leuvenlaan 4, Postbus 80.195, 3508 TD Utrecht, THE NETHERLANDS}
\end{center}

\vspace{0.2cm}

\vspace{0.3cm}

\begin{center}
ABSTRACT
\end{center}
\hspace{0.3cm} In this paper we consider a massless scalar field,
with a possible coupling $\xi$ to the Ricci scalar in a $D$
dimensional FLRW spacetime with a constant deceleration parameter
$q=\epsilon-1$, $\epsilon=-{\dot{H}}/{H^2}$. Correlation functions
for the Bunch-Davies vacuum of such a theory have long been known
to be infrared divergent for a wide range of values of $\epsilon$.
We resolve these divergences by explicitly matching the spacetime
under consideration to a spacetime without infrared divergencies.
Such a procedure ensures that all correlation functions with
respect to the vacuum in the spacetime of interest are infrared
finite. In this newly defined vacuum we construct the coincidence
limit of the propagator and as an example calculate the
expectation value of the stress energy tensor. We find that this
approach gives both in the ultraviolet and in the infrared
satisfactory results. Moreover, we find that, unless the effective
mass due to the coupling to the Ricci scalar $\xi R$ is negative,
quantum contributions to the energy density always dilute away
faster, or just as fast, as the background energy density.
Therefore, quantum backreaction is insignificant at the one loop
order, unless $\xi R$ is negative. Finally we compare this
approach with known results where the infrared is regulated by
placing the Universe in a finite box. In an accelerating universe,
the results are qualitatively the same, provided one identifies
the size of the Universe with the physical Hubble radius at the
time of the matching. In a decelerating universe however, the two
schemes give different late time behavior for the quantum stress
energy tensor. This happens because in this case the length scale
at which one regulates the infrared becomes sub-Hubble at late
times.
 \vspace{0.3cm}

\begin{flushleft}
\end{flushleft}

\vspace{0.1cm}

\begin{flushleft}
$^{*}$ T.M.Janssen@uu.nl, $^{\dagger}$ T.Prokopec@uu.nl
\end{flushleft}

\end{titlepage}
\section{Introduction}
It is by now well established that on the largest scales, the
universe is well described by a homogeneous, isotropic and
spatially flat metric, given by~\cite{Mukhanov:2005sc}
\begin{equation}\label{metric}
    g_{\mu\nu}dx^\mu dx^\nu = -dt^2+a(t)^2 d\vec{x}\cdot d\vec{x}.
\end{equation}
Here $a$ is the scale factor, and the derivatives on $a$ define
the Hubble parameter, $H$ and the deceleration parameter $q$ (in
this paper we shall use the equivalent parameter $\epsilon$)
\begin{equation}
    H\equiv \f{\dot{a}}{a}\qquad;\qquad q \equiv
    -1-\f{\dot{H}}{H^2}\equiv -1+\epsilon
\,.
\label{epsilon:def}
\end{equation}
If $\epsilon>1$ the expansion of the universe is decelerating, if
$\epsilon<1$ it is accelerating and if $\epsilon=1$ it is
curvature dominated. In a typical cosmological model, we couple
this geometry to a certain perfect fluid, with energy density
$\rho_b=\rho_b(t)$ and pressure $p_b=p_b(t)$. The dynamics if the
scale factor are then governed by the Friedmann equations
\begin{equation}\label{Fried}
    \begin{split}
    H^2&=\f{8\pi G_N}{3}\rho_b\\
    \dot{H}&=-4\pi G_N (\rho_b+p_b)
\,,
    \end{split}
\end{equation}
where $G_N$ indicates Newton's constant and for simplicity we
assumed flat spatial sections. This is consistent with the current
cosmological measurements~\cite{Komatsu:2008hk}, which give a very
large lower limit on the radius of curvature $R_c$ of the
Universe, $R_c\geq 22h^{-1}~{\rm Gpc}$ ($R_c\geq 33h^{-1}~{\rm
Gpc}$) for a universe with positively (negatively) curved spatial
sections, where $h=0.705\pm 0.013$ is the rescaled Hubble
parameter (in units of $100~{\rm km/s/Mpc}$). If we now assume
that the cosmological fluid obeys an equation of state
\begin{equation}\label{eq_of_state}
    p_b=w_b\rho_b,
\end{equation}
 with $w$ constant
we immediately find that in such a case
\begin{equation}
    \epsilon = \f{3}{2}(1+w_b) = \mathrm{constant}.
\label{epsilon:constant}
\end{equation}
The equation of state (\ref{eq_of_state}) is in practice obeyed
for many types of fluids. For example radiation corresponds to
$w_b=1/3$ and $\epsilon=2$, non-relativistic matter has $w_b=0$
and $\epsilon=3/2$, the cosmological constant has $w_b=-1$ and
$\epsilon=0$. If $\epsilon$ is a constant, we can find the Hubble
parameter and the scale factor as a function of time
\begin{equation}\label{H_and_a}
H(t) = \frac{H_0}{1+\epsilon H_0 t}\qquad;\qquad
a(t)=\Big(1+\epsilon H_0 t\Big)^{\frac{1}{\epsilon}},
\end{equation}
with $H_0$ a constant.
\\
The behavior of a massless scalar field with a possible coupling
to the Ricci scalar on such a background geometry has been
extensively studied in the
literature~\cite{BD}\cite{Vilenkin:1982wt}\cite{FP}\cite{AV}. An
interesting observation is that the kinetic operator of, for
example, the graviton can be related to the kinetic operator of
such a massless scalar field. Therefore the propagator of the
graviton can, apart from some tensorial structure, be written in
terms of propagators of massless scalar
fields~\cite{Grish}\cite{JP2}. Similar observations can be made
for vector fields~\cite{Tsamis:2006gj} or anti-symmetric tensor
fields \cite{Janssen:2007yu}. Therefore an understanding of the
physics of the massless scalar field can be translated to those
fields as well.\\
In particular many studies concern in particular the special limit
of $\epsilon\rightarrow 0$ (de Sitter) ~\cite{Linde:1982uu}.
Because of the nature of the background space-time, the scalar
field has non-trivial quantum properties. In particular, due to
the presence of a horizon, there is particle production. The
particles are created mostly in the infrared, and it is this
effect that leads for example to the creation of primordial
density fluctuations. Also it is the basis for many works of the
quantum backreaction on the background
space-time~\cite{Tsamis:1996qm}\cite{Tsamis:1996qq}\cite{Abramo:1996gd}\cite{Abramo:1997hu}\cite{Abramo:1999wd}\cite{Mukhanov:1996ak}\cite{Losic:2006ht}.
However, it was long ago realized that when one chooses the vacuum
to contain only purely positive frequency modes (the so called
Bunch-Davies vacuum), the expectation of the two point correlator
for this vacuum diverges in the infrared, for a large -- and
physically relevant -- range of $\epsilon$~\cite{FP}. In
particular, if the massless scalar field is minimally coupled,
there are infrared divergencies for all $\epsilon\leq 3/2$. The
presence of these divergencies implies that in the infrared, the
Bunch-Davies vacuum does not describe a physically sensible state.
One possible solution to this problem is to assume that the
spacetime manifold is spatially compact, for example a torus
$T^{D-1}$. This approach effectively introduces an infrared
cut-off, when the sum over the modes is approximated by an
integral~\cite{TW3}. In Ref.~\cite{Janssen:2008px} the propagator
is explicitly constructed using this regularization
for any constant $\epsilon$ space.\\
A different approach is to choose the mode
functions such that the super-horizon modes are less singular than
they would have been in the Bunch-Davies vacuum
~\cite{AV,Vilenkin:1982wt}. Because only the super-horizon modes
change there would be no effect on the Hadamard short distance
behavior of the propagator. The time dependence of the mode
functions is determined by the scalar field equation but their
initial values and the initial values of their first time
derivatives can be freely specified. Thus we can choose the
initial values for the super-Hubble infrared modes to be in some
infrared finite state. Such a choice would ensure then that there
would be no infrared divergence, either initially or at
any later time~\cite{Fulling:1978ht}.\\
In this paper however, we propose a third regularization of the
infrared. Instead of splitting the ultraviolet and the infrared
sector, one could also consider the matching between the
Bunch-Davies vacuum in an infrared safe space-time and the
space-time one wishes to study for all modes. Also in this case
the initial state does not lead to infrared divergences and thus
also the final state will be safe. In this approach the
ultraviolet mode functions will differ from the ultraviolet
Bunch-Davies mode functions, but the ultraviolet divergent
structure will not change. The advantage of this approach over the
previous two is that one can quite easily envision something like
this to be realized in nature. For example if inflation was
preceded by a radiation dominated epoch, we essentially have the
scenario described above, if the field was in its Bunch-Davies
vacuum during the radiation epoch and if the transition between
the two geometries is fast enough. A disadvantage is that the
sudden matching between the two geometries causes the Ricci scalar
to change discontinuously. This causes a burst of particle
creation, but we shall see that this does not significantly
influence our main results. \\
There are other regularization schemes proposed in literature.
Here we mention a recent proposal by Parker~\cite{Parker:2007ni},
where it was argued that the infrared should be regulated by
subtracting the adiabatic vacuum contribution. The proposal has
been worked out to some detail in Ref.~\cite{Agullo:2009vq}. We
feel that that this scheme is not well motivated, since the
infrared sector of constant $\epsilon$ spaces strongly breaks
adiabaticity, which is the assumption made when constructing
adiabatic vacua. Yet another possibility is to subtract the vacuum
of comoving observers~\cite{Agullo:2008ka}. This procedure works
well for de Sitter space~\cite{Agullo:2008ka}, but it remains to
be seen whether such a procedure can be successfully implemented
in more general FLRW spacetimes.

In this paper we want to study the role of the quantum infrared
fluctuations generated by the expansion of the Universe for a
massless scalar field. In particular we would like to understand
whether the energy density due to these fluctuations could ever
become relevant for the evolution of the universe. For a proper
understanding of such a question, it is important to compare
different regularization schemes for the infrared. In particular
we expect that in an accelerating space-time, results will not
depend that much on the regularization chosen, since the details
of any such scheme will grow quickly to super-Hubble scales. The
reason is that in accelerating space-times, physical length scales
grow faster then the Hubble radius. In a decelerating space-time
however, the opposite is true. Specific cases need to be studied
to understand this issue in detail.

In section \ref{s_scalar} we discuss the construction of the
scalar field mode functions and its infrared properties. In
section \ref{s_matching} we construct the new vacuum by matching
it to an infrared finite Bunch-Davies vacuum. The associated
propagator is constructed in section \ref{s_prop}. We discuss some
properties of the stress energy tensor in section \ref{s_stress}
and we calculate the ultraviolet contribution to the stress energy
tensor in section \ref{reormalization} and consider its
renormalization. In section \ref{The equation of state of the
quantum fluid} we consider the equation of state of the quantum
fluid and study the scaling of the quantum energy density. In
section~\ref{s_cutoff} we compare our results with the results
obtained in Ref.~\cite{Janssen:2008px} using a cut-off procedure
and finally in section \ref{Summary and discussion} we summarize
our results and conclude.

\section{A Massless scalar field}\label{s_scalar}

 We wish to study the behavior of a massless scalar field $\phi$ in
the geometry (\ref{metric}). The action in $D$ dimensions is given
by
\begin{equation}\label{action}
    S=\f{1}{2}\int d^Dx\sqrt{-g}\phi\big(\Box-\xi R\big)\phi,
\end{equation}
where $\xi$ is a constant coupling of the scalar field to the Ricci
scalar $R$. The equations of motion for $\phi$ are
\begin{equation}\label{EOM}
    (\Box-\xi R)\phi=0,
\end{equation}
where in our background metric the scalar d'Alembertian reads,
\begin{equation}
    \Box = -\f{\partial^2}{\partial t^2}
    +\f{1}{a^2}\f{\partial ^2}{\partial \vec{x}^{\,2}}
  -(D-1)H\f{\partial }{\partial t}
\,.
\end{equation}
 Now we use (spatially) Fourier transformed field
\begin{equation}
    \tilde{\phi}(t,k) = \int d^{D-1}x e^{-i\vec{k}\cdot\vec{x}}\phi(x)
\end{equation}
to obtain the equations of motion for $\tilde{\phi}$
\begin{equation}
    \ddot{\tilde{\phi}}+(D-1)H\dot{\tilde{\phi}}
    +\f{k^2}{a^2}\tilde{\phi}+\xi R\tilde{\phi}=0
\,\quad (k=\|\vec k\|\,)
\,,
\end{equation}
where $\dot{\tilde{\phi}}=\partial\tilde \phi/\partial t$.
We expand our solution in (time independent) creation and
annihilation operators $b^{\dagger}$ and $b$
\begin{equation}
    \tilde{\phi}(t,\vec k\,) = \psi(t,k)b(\vec k\,)
    +\psi^{*}(t,k)b^{\dagger}(-\vec k\,)
    \,,
\end{equation}
where the spatial homogeneity of the mode functions $\psi(t,k)$
implies that they are a function of the modulus of $\vec k$, and
$b(\vec k\,)$ is the annihilation operator that annihilates the
vacuum $|\Omega\rangle$, $b(\vec k\,)|\Omega\rangle=0$. The
normalization is fixed by requiring the canonical commutation
relations
\begin{equation}
    \begin{split}
        \Big[b(\vec k\,),b^\dagger(\vec k'\,)\Big]
          &=(2\pi)^{D-1}\delta^{D-1}(\vec k-\vec k'\,)\\
        \Big[\tilde{\phi}(t,\vec k\,),
             a(t)^{D-1}\dot{\tilde{\phi}}^\dagger(t,\vec k'\,)\Big]&
                  =i(2\pi)^{D-1}\delta^{D-1}(\vec k-\vec k'\,)
                  \,,
        \end{split}
\end{equation}
which implies that the Wronskian of the mode functions $\psi$ is
given by
\begin{equation}\label{wronskian}
    \psi(t,k)\dot{\psi}^{*}(t,k)-\psi^{*}(t,k)\dot{\psi}(t,k)=i a^{1-D}
\,.
\end{equation}
The mode functions $\psi$ obey the equation
\begin{equation}
    \Big(\f{d^2}{dt^2}+H\f{d}{dt}+\f{k^2}{a^2}
    +\f{D-2\epsilon}{4}\big(2-D+4\xi(D-1)\big)H^2\Big)\Big(a^{\f{D}{2}-1}\psi\Big)=0,
\end{equation}
where we used that in the constant $\epsilon$ geometry under
consideration~\cite{Janssen:2007ht}
\begin{equation}
    R=(D-1)(D-2\epsilon)H^2.
\end{equation}
We write $z=\f{k}{(1-\epsilon)Ha}$. If we assume that $\epsilon$
is constant as in~(\ref{epsilon:constant}), we obtain
\begin{equation}\label{EOM2}
    \Big(z^2\f{d^2}{dz^2}+z^2+\f{1}{4}-\nu^2
    \Big)\Big(a^{\f{D}{2}-1}\psi\Big)=0,
\end{equation}
where
\begin{eqnarray}\label{nu}
 \nu^2 &=& \Big(\f{D-1-\epsilon}{2(1-\epsilon)}\Big)^2
       - \f{(D-1)(D-2\epsilon)}{(1-\epsilon)^2}\xi
\nonumber\\
       &=& \f14
       - \f{(D-2\epsilon)}{(1-\epsilon)^2}\Big((D-1)\xi-\f{D\!-\!2}{4}\,\Big)
\,,
\end{eqnarray}
such that $\nu=1/2$ for a conformally coupled scalar, for which
$\xi=\xi_c\equiv (D-2)/[4(D-1)]$, or when the universe is
radiation dominated, such that $\epsilon=D/2$.
Equation~(\ref{EOM2}) can be easily related to Bessel's equation
and we find the solution to be
\begin{equation}\label{mode_sol}
    \begin{split}
        \psi(t,k) &= a^{1-\f{D}{2}}\Big(\alpha(k) u(t,k) + \beta(k)
        u^{*}(t,k)\Big)\\
        u(t,k)&=\sqrt{\f{\pi}{4(1-\epsilon)H
        a}}H^{(1)}_\nu\Big(\f{k}{(1-\epsilon)H a}\Big)
        =\sqrt{\f{\pi z}{4k}}H^{(1)}_\nu(z)
        \,,
    \end{split}
\end{equation}
where $H^{(1)}_\nu$ denotes the first Hankel function. The
normalization is chosen such that the Wronskian condition
(\ref{wronskian}) implies that
\begin{equation}\label{alpha_beta_wronsk}
    |\alpha|^2-|\beta|^2=1
    \,.
\end{equation}
The time-ordered (Feynman)
propagator for a given vacuum state $|\Omega\rangle$ is now given by
\begin{equation}\label{propdef}
\begin{split}
    i\Delta(x;x')&\equiv
    \langle\Omega\big|\theta(t-t')\phi(x)\phi(x')+\theta(t'-t)\phi(x')\phi(x)\big|\Omega\rangle\\
    &=\int\f{d^{D-1}k}{(2\pi)^{D-1}}e^{i\vec{k}\cdot(\vec{x}-\vec{x'})}
    \Big(\theta(t-t')\psi(t)\psi^{*}(t')+\theta(t'-t)\psi(t')\psi^{*}(t)\Big)
    \,,
    \end{split}
\end{equation}

\begin{figure}
\begin{center}
\includegraphics[width=6in]{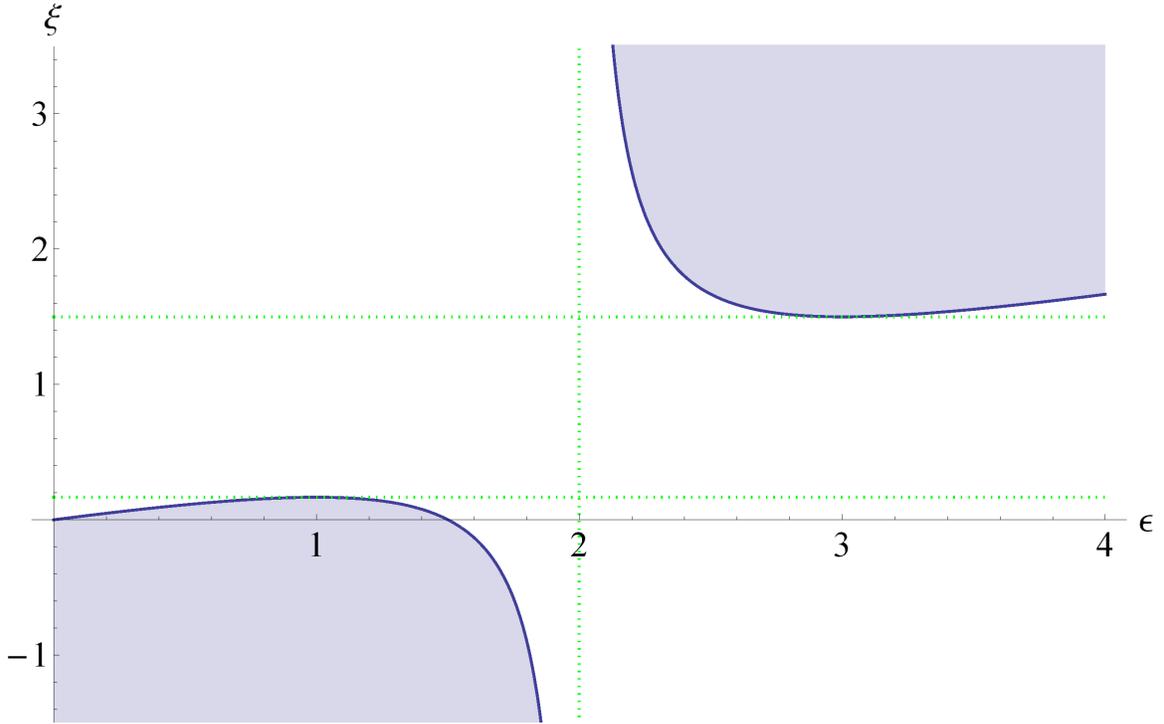}
\caption{The shaded regions indicate the regions where
$\nu\geq\f{3}{2}$, with $\nu$ given in (\ref{nu}) and we choose
$D=4$. When this requirement is met, the coincident propagator
(\ref{IRdiv:BD}) is infrared divergent. The regions are bounded by
the curve $\xi=\f{\epsilon}{6}\f{3-2\epsilon}{2-\epsilon}$ and the
dotted asymptotes are $\epsilon=2$, $\xi=1/6$ and $\xi=3/2$.
}\label{fig0}
\end{center}
\end{figure}

The vacuum $|\Omega\rangle$ is defined by $b(\vec
k\,)|\Omega\rangle=0$. This vacuum state is clearly not unique,
since the commorients $\alpha$ and $\beta$ are not uniquely
determined by (\ref{alpha_beta_wronsk}). The particular choice

\begin{equation}
    \alpha =1\qquad;\qquad\beta=0
    \,.
\label{BD vacuum}
\end{equation}
corresponds to a field $\phi$ in a state $|\Omega\rangle$  known
as the Bunch-Davies vacuum~\cite{BD}. This choice minimizes the
energy in the ultraviolet modes and is therefore often considered
as a reasonable choice for the vacuum. By considering the small
argument limit of the Bessel function (see Eq.~(8.402) of
Ref.~\cite{GR})
\begin{equation}
    J_\nu(z)=\sum_{n=0}^{\infty}\f{(-1)^n}{n!\Gamma(\nu+n+1)}\Big(\f{z}{2}\Big)^{\nu+2n}
\end{equation}
and using that (see Eqs.~(8.403) and~(8.405.1) of Ref.~\cite{GR})
\begin{equation}
    H^{(1)}_\nu(z)=\f{i}{\sin(\pi\nu)}\Big(e^{-i\pi\nu}J_\nu(z)-J_{-\nu}(z)\Big)
\,,
\end{equation}
we find that the small $k$ contribution to
the integrand of (\ref{propdef}) in the case of the Bunch-Davies
vacuum behaves as
\begin{equation}
   i\Delta(x;x)_{\rm BD}\propto\int k^{D-2-2\nu}(1+\mathcal{O}(k^2)) dk
\,,
\label{IRdiv:BD}
\end{equation}
where we took $\nu>0$ in Eq.~(\ref{nu}). We thus see that the
propagator of the Bunch-Davies vacuum is infrared divergent for
all~\cite{FP,Janssen:2008px}
\begin{equation}\label{IR_req}
\nu\geq \f{D-1}{2}.
\end{equation}
For example if $\xi=0$ and $D=4$ (we do not need dimensional
regularization in the infrared) this implies that there is an
infrared divergence for all $\epsilon\leq 3/2$. From
(\ref{epsilon:constant}) we find that this implies that the
pressure of the fluid driving the Universe's expansion is
negative. In the more general case, where $\xi\neq 0$, we refer to
figure (\ref{fig0}), where we plotted (in D=4) the regions in the
$\epsilon-\xi$ plane where the requirement (\ref{IR_req}) is met.
We can qualitatively understand this picture, by realizing that
the coupling $\xi$ acts as an effective, time-dependent mass term
for the scalar field proportional to $\xi R =6\xi(2-\epsilon)H^2$.
Thus for $\epsilon>2$. Thus we have an infrared divergence if this
mass term is 'sufficiently' less then 1/6 (where the precise
meaning of 'sufficiently' follows from (\ref{IR_req}).

\section{Matching}\label{s_matching}

To resolve the infrared divergences, we shall consider a geometry
where $\nu=1/2$ at times $t<\hat{t}$, while $\nu$ is an arbitrary
half integer for $t>\hat{t}$. Notice that $\nu=1/2$ corresponds to
a radiation dominated universe ($\epsilon=D/2$) if $\xi=0$, or a
universe whose expansion is driven by a conformally coupled scalar
($\xi=\f{D-2}{4(D-1)}$). The restriction that $\nu$ after the
matching should be half integer comes from technical
considerations. The calculation simplifies enormously in this
case, since the mode functions can now be written as finite sums.
Moreover we shall find that in the end of the calculation we can
resum these finite sums and analytically extent the result,
obtaining thus an answer, which then will be
valid for arbitrary $\nu$.\\
We match these two geometries, such that the scale factor $a(t)$
and the Hubble parameter $H(t)$ are continuous at the matching.
This requires that also the mode functions and their first
derivative need to be continuous at the matching~\cite{TW2}. It is
clear from (\ref{IR_req}) that before the matching, in the
$\nu=1/2$ spacetime,  we can freely choose the Bunch-Davies
vacuum. Since there is no infrared
divergence initially, it will not be there at any later time.\\
Quantities before the matching we indicate with a subscript $0$,
while quantities after the matching do not have a subscript. Thus
for $t<\hat{t}$ we have $\epsilon_0$ and for $t>\hat{t}$ we have
$\epsilon$. The mode functions (\ref{mode_sol}) are written in
terms of the following solutions
\begin{equation}
    \begin{split}
    u_0(t<\hat{t},k) &= -\f{i}{\sqrt{2k}} e^{\f{ik}{(1-\epsilon_0)H
    a}}\\
    u(t>\hat{t},k) &= \sqrt{\f{\pi}{4(1-\epsilon)H
    a}}H^{(1)}_\nu\Big(\f{k}{(1-\epsilon)H a}\Big)\\
    &=\f{1}{\sqrt{2k}}(i)^{-\nu-\f{1}{2}}e^{\f{ik}{(1-\epsilon)H
    a}}\sum_{n=0}^{\nu-\f{1}{2}}\f{\Gamma(\nu+\f{1}{2}+n)}{n!\Gamma(\nu+\f{1}{2}-n)}\Big(\f{-2ik}{(1-\epsilon)H
    a}\Big)^{-n},
    \end{split}
\end{equation}
where the last line follows from the fact that $\nu$ is a half
integer~\cite{GR}. We choose the initial vacuum to be Bunch-Davies,
and therefore the mode functions are
\begin{equation}
    \begin{split}
    \psi_0(t<\hat{t},k)&=a^{1-\f{D}{2}}u_0(t,k)\\
    \psi(t>\hat{t},k)&=a^{1-\f{D}{2}}\Big(\alpha(k)u(t,k)+\beta(k) u^{*}(t,k)\Big).
    \end{split}
\end{equation}
We shall now fix the coefficients $\alpha$ and $\beta$ by
requiring that at the matching $a$ and $H$ are continuous and thus
$\psi$ and $\f{d}{dt}\psi$ should be continuous
\begin{equation}
    \begin{split}
    \psi_0(\hat{t},k)&=\psi(\hat{t},k)\\
    \f{d}{dt}\psi_0(t,k)\Big|_{t=\hat{t}}&=\f{d}{dt}\psi(t,k)\Big|_{t=\hat{t}}.
    \end{split}
\end{equation}
These conditions imply that
\begin{equation}\label{alphabeta}
    \begin{split}
        \alpha&=i
        a\Big(u^{*}\dot{u}_0-\dot{u}^{*}u_0\Big)\Big|_{t=\hat{t}}\\
        &=i^{\nu+\f12}e^{-\f{ik}{p}\f{\epsilon-\epsilon_0}{(1-\epsilon)(1-\epsilon_0)}}
        \sum_{n=0}^{\nu-\f{1}{2}}\Big(-i-\f{(1-\epsilon)np}{2k}\Big)
        \f{\Gamma(\nu+\f{1}{2}+n)}{n!\Gamma(\nu+\f{1}{2}-n)}\Big(\f{2ik}{(1-\epsilon)p}\Big)^{-n}
\\
         \beta&=i
        a\Big(\dot{u} u_0-u\dot{u}_0\Big)\Big|_{t=\hat{t}}\\
        &=i^{-(\nu+\f12)}e^{-\f{ik}{p}\f{\epsilon+\epsilon_0-2}{(1-\epsilon)(1-\epsilon_0)}}
        \sum_{n=0}^{\nu-\f{1}{2}}\Big(\f{(1-\epsilon)np}{2k}\Big)
        \f{\Gamma(\nu+\f{1}{2}+n)}{n!\Gamma(\nu+\f{1}{2}-n)}\Big(\f{-2ik}{(1-\epsilon)p}\Big)^{-n}
\,,
    \end{split}
\end{equation}
where we defined $p \equiv a(\hat{t}) H(\hat{t})$. Notice that the
series expansion used for the Hankel functions is also valid also
for negative values of the argument, so we do not run into any
problems due to the branch cut of the Hankel function for negative
real arguments.

\section{The Coincidence Propagator}\label{s_prop}
Now we shall calculate the propagator, using the state
$|\Omega\rangle$ defined by~(\ref{alphabeta}). In particular, we
shall only calculate the propagator at coincidence. As we shall
see this is sufficient for our present purpose, calculating the
one loop corrected stress-energy tensor. The coincidence
propagator is given by~\cite{Janssen:2008px}
\begin{equation}\label{propcoinc}
    i\Delta(x;x)=\f{1}{2^{D-2}\pi^{\f{D-1}{2}}\Gamma(\f{D-1}{2})}
    \int dk
    k^{D-2}|\psi(t,k)|^2
    \,.
\end{equation}
Using the expressions from the previous section we obtain
\begin{equation}\label{int}
\begin{split}
    i\Delta(x;x) &=
    \f{a^{2-D}}{(4\pi)^{\f{D-1}{2}}\Gamma(\f{D-1}{2})}\int dk k^{D-3}
    \\
    &\sum_{q,r,m,n=0}^{\nu-\f{1}{2}}
    \Bigg\{\f{\Gamma(\nu+\f{1}{2}+n)\Gamma(\nu+\f{1}{2}+m)\Gamma(\nu+\f{1}{2}+q)\Gamma(\nu+\f{1}{2}+r)}
    {\Gamma[\nu+\f{1}{2}-n)\Gamma(\nu+\f{1}{2}-m)\Gamma(\nu+\f{1}{2}-q)\Gamma(\nu+\f{1}{2}-r)}\\
    &\Bigg[\bigg(1+\f{i}{2}(1-\epsilon)(q-n)\f{p}{k}+\f{1}{2}(1-\epsilon)^2nq\f{p^2}{k^2}\bigg)(-1)^{-q-m}\\
    &+e^{\f{2ik}{(1-\epsilon)Ha}(1-\f{aH}{p})}\Big(-\f{i}{2}(1-\epsilon)q\f{p}{k}-\f{1}{4}(1-\epsilon)^2 qn\f{p^2}{k^2}\Big)(-1)^{-r-m}\\
    &+e^{\f{-2ik}{(1-\epsilon)Ha}(1-\f{aH}{p})}\Big(\f{i}{2}(1-\epsilon)q\f{p}{k}-\f{1}{4}(1-\epsilon)^2 qn\f{p^2}{k^2}\Big)(-1)^{-q-n}\Bigg]\\
    &\Big(\f{2ik}{(1-\epsilon)p}\Big)^{-n-q}\Big(\f{2ik}{(1-\epsilon)aH}\Big)^{-m-r}\f{1}{n!m!q!r!}\Bigg\}
\,.
    \end{split}
\end{equation}
One important property of this expression is that the integral
converges in the infrared, or $k\rightarrow 0$. Of course this was
expected by construction, but it can be shown explicitly. For
example the quadruple sum in (\ref{int}) evaluates in some
explicit cases to
\begin{equation}
    \begin{split}
        \sum_{q,r,m,n=0}^{\nu-\f{1}{2}}\Big(\ldots\Big)_{\nu=3/2}&=\Big(\f{aH}{p}\Big)^2\Big(\f{2}{3}+\f{1}{3}\big(\f{p}{aH}\big)^3\Big)^2+\mathcal{O}(k)\\
        \sum_{q,r,m,n=0}^{\nu-\f{1}{2}}\Big(\ldots\Big)_{\nu=5/2}&=\Big(\f{aH}{p}\Big)^4\Big(\f{3}{5}+\f{2}{5}\big(\f{p}{aH}\big)^5\Big)^2+\mathcal{O}(k)\\
        \sum_{q,r,m,n=0}^{\nu-\f{1}{2}}\Big(\ldots\Big)_{\nu=7/2}&=\Big(\f{aH}{p}\Big)^6\Big(\f{4}{7}+\f{3}{7}\big(\f{p}{aH}\big)^7\Big)^2+\mathcal{O}(k).
    \end{split}
\end{equation}

\subsection{The Ultraviolet}\label{sec_UV}

 As long as $aH\neq p$, all UV divergencies can be easily seen to
come only from the first line in square brackets in
Eq.~(\ref{int}). In fact the second line has three types of
contributions: UV divergent terms, IR divergent terms and a term
that is logarithmically divergent in both the UV and the IR. Now
we know that the IR divergent terms will cancel against the IR
divergent terms coming from the second and third lines in the
square brackets. So we shall simply drop all IR divergent terms
that we encounter (except for the logarithmic divergence). By
explicitly calculating the first 3 terms of the sums, we find that
the UV divergent contributions are
\begin{equation}
    i\Delta(x;x)_{UV}
    =\f{a^{2-D}}{(4\pi)^{\f{D-1}{2}}\Gamma(\f{D-1}{2})}\int\Big(k^{D-3}-\f{a^2
    H^2}{8}k^{D-5}(1-\epsilon)^2(1-4\nu^2)\Big)dk
\label{UV:divergence}
\end{equation}
The first (quadratically divergent) term is a scaleless integral,
which we can automatically subtract in dimensional
regularization~\cite{HV}\cite{BG}. The second contribution is
logarithmically divergent. Since we know that the infrared
divergent part should drop out in the final answer, we write the
integral as
\begin{equation}\label{intsplit}
    \int\rightarrow\int_0^{k_0}+\int_{k_0}^{\infty}
\end{equation}
and we drop the first, IR divergent, integral. Now we obtain

\begin{equation}\label{propUV}
    i\Delta(x;x)_{UV}=\f{H^2}{8\pi^2}(1-\epsilon)^2\Big(\nu^2-\f14\Big)
    \Big[-\f{\mu^{D-4}}{D-4}-\f{\gamma_E}{2}+1
    +\f12\ln\Big(\f{\pi\mu^2a^2}{k_0^2}\Big)+\mathcal{O}(D-4)\Big]
\,,
\end{equation}
where one should note that the prefactor
$(1-\epsilon)^2(\nu^2-\f{1}{4})$ is still $D$-dependent, as can be
seen from  (\ref{nu}),
 and we introduced a
renormalization scale $\mu$. The $k_0$ dependence of
(\ref{propUV}) should -- and does --  drop out when we add the IR
contribution. Notice that $\Delta(x;x)_{UV}$ is exactly the same
as the (divergent) UV contribution one would have obtained using
the Bunch-Davies vacuum~\cite{Janssen:2008px}. This was to be
expected, since the ultraviolet of the theory is only sensitive to
local physics and therefore does not 'see' the matching.

\subsection{The Infrared}

We now focus on the second line in square brackets of (\ref{int})
(the third line is simply the complex conjugate of the second
line). As long as $aH \neq p$, the contributions are UV finite, so
we can put $D=4$. We shall comment below on the special case
$aH=p$. The integrals are of the form
\begin{equation}
    \int k^{-n}e^{i\alpha k}  dk,
\nonumber
\end{equation}
where $n$ is a positive integer and $\alpha<0$. We split the
integral again in two ranges as in (\ref{intsplit}) and take the
limit $k_0\rightarrow 0$. The lower integral will be divergent,
but these divergences exactly cancel against the infrared
divergences coming from the second line of (\ref{int}). The upper
integral is given by ($\Im\big[\alpha\big]>0$)
\begin{equation}\label{generalint0}
    \int_{k_0}^{\infty} k^{-n} e^{i\alpha k}dk
    = k_0^{1-n}E_n(-i\alpha k_0)
\,,\end{equation}
where $n$ is an integer and
\begin{equation}
 E_p(z) = \int_{1}^\infty dt\f{{\rm e}^{-z t}}{t^p}
\nonumber
\end{equation}
denotes the exponential integral. For integer index, the
exponential integral has the following expansion around zero
\begin{equation}
    E_n(z)=\f{(-z)^{n-1}}{\Gamma(n)}\Big(\psi(n)-\ln(z)\Big)-\sum_{k=0\,,\,\,k\neq
    n-1}^{\infty} \f{(-z)^k}{(k-n+1)k!}
\,,
\end{equation}
where $\psi(z)=(d/dz)\ln[\Gamma(z)]$ denotes the digamma function.
Since we know that all negative powers of $k_0$ cancel, we obtain
the following leading order result, which is obtained by expanding
(\ref{generalint0}) around $k_0=0$,
 \begin{equation}\label{generalint}
    \int_{k_0}^{\infty} k^{-n}e^{\pm i\alpha k}dk
    = -\f{(\pm i\alpha)^{n-1}}{\Gamma(n)}\left[\ln(\mp i\alpha k_0)-\psi(n)\right]
    +({\rm IR-div})+{\cal O}(k_0)
\,,
\end{equation}
where $\psi(z)=(d/dz)\ln[\Gamma(z)]$ denotes the digamma function.

We use this expression to evaluate the third and fourth line in
(\ref{int}) and obtain
\begin{equation}\label{propIR}
    \begin{split}
i\Delta(x;x)_{IR} =&
\f{H^2(1-\epsilon)^2}{4\pi^2}\sum_{m,n,q,r=0}^{\nu-\f{1}{2}}
\f{\Gamma(\nu+\f{1}{2}+n)\Gamma(\nu+\f{1}{2}+m)\Gamma(\nu+\f{1}{2}+q)\Gamma(\nu+\f{1}{2}+r)}
{\Gamma(\nu+\f{1}{2}-n)\Gamma(\nu+\f{1}{2}-m)\Gamma(\nu+\f{1}{2}-q)\Gamma(\nu+\f{1}{2}-r)}
\\
&\times\Bigg[n(1\!-\!\zeta)\f{\ln(|\alpha|k_0)-\psi(m+n+q+r+1)}{\Gamma(m+n+q+r+1)}
       -\f{\ln(|\alpha|k_0)-\psi(m+n+q+r)}{\Gamma(m+n+q+r)}
\Bigg]
\\
&\times\f{q}{2}\f{\zeta^2}{1-\zeta}\Big(-(1-\zeta)\Big)^{n+q}
\Big(\f{1}{\zeta}-1\Big)^{m+r}\f{1}{n!m!q!r!} \,,
\end{split}
\end{equation}
where $\alpha=\f{2}{(1-\epsilon)aH}(1-\zeta^{-1})$ and $\zeta=p/(aH)$.

 Quite remarkably, the sums in Eq.~(\ref{propIR}) can be
(almost completely) performed, resulting in:
\begin{equation}\label{propIR:2}
    \begin{split}
i\Delta(x;x)_{IR} =&
\f{H^2(1-\epsilon)^2}{8\pi^2}\Big(\nu^2-\frac14\Big)
\bigg[\sum_{n=1}^{\nu-3/2}\f{1}{2n \zeta^{2n}}
   + \ln(|\alpha|k_0) -\f12 c_\nu - 1 +\f12\gamma_E
  + \zeta + \sum_{n=2}^{\nu+1/2}\f{\zeta^{2n}}{2n}
\bigg]
\,,
\end{split}
\end{equation}
where $c_\nu$ is a slowly varying function of $\nu$~\footnote{We
were unable to evaluate the constant $c_\nu$. The first few values
are given by quite simple expressions. For
$\nu=\{3/2,5/2,7/2,9/2,11/2\}$ we have
$c_\nu=\psi(2\nu-2)+\{1/2,1/3,1/2,2/3,137/168\}.$}. Its precise
value is not important, as it can be absorbed in the  by a finite
counterterm.

\subsection{The Full Propagator at coincidence}
\label{The Full Propagator at coincidence}

The full propagator at coincidence is given by the sum of the
infrared and ultraviolet contributions.
 Adding the infrared~(\ref{propIR:2})
 and ultraviolet~(\ref{propUV}) contributions
we obtain,
\begin{equation}\label{prop:full}
    \begin{split}
i\Delta(x;x) =&
\f{H^2(1-\epsilon)^2}{8\pi^2}\Big(\nu^2-\frac14\Big)
\bigg[-\f{\mu^{D-4}}{D-4}-\f{c_\nu}{2}
+\f12\ln\bigg(\f{4\pi\mu^2}{(1-\epsilon)^2H^2}\Big(1-\f1\zeta\Big)^2\bigg)
\\
&  + \zeta -\f{1}{2}\zeta^2+
\sum_{n=1}^{\nu-3/2}\frac{\zeta^{-2n}}{2n} +
\sum_{n=1}^{\nu+1/2}\frac{\zeta^{2n}}{2n} \bigg] \qquad \Big(
\zeta= \frac{\hat{H}{\hat a}}{Ha}\Big) \,.
\end{split}
\end{equation}
We see indeed that, as promised, the two logarithms of $k_0$
coming from the infrared and the ultraviolet have canceled.
From the result~(\ref{prop:full}) we see
immediately that when $|\zeta|\ll 1$ the leading order contribution to
$i\Delta(x;x)$
 is of the form, $\propto H^2\zeta^{3-2\nu}$, while when
$|\zeta|\gg 1$ the leading order contribution goes as $\propto
H^2\zeta^{2\nu+1}$.

The two series in~(\ref{prop:full}) can be resummed, resulting in:
\footnote{{\tt Mathematica} represents the answer
in terms of the Lerch transedent,
${\tt LerchPhi}[z,s,\alpha]$, which is the following
generalization of the Riemann $\zeta$ function,
${\tt LerchPhi}[z,s,\alpha]=\sum_{n=0}^\infty z^n/(n+\alpha)^s$.
For our purpose it is convenient to express the Lerch
transedent in terms of the Gauss' hypergeometric function,
\[
 {\tt LerchPhi}[z,1,\alpha] = \frac{{}_2F_1(1,\alpha;\alpha+1;z)}{\alpha}
\,.
\]}
\begin{eqnarray}
i\Delta(x;x) \!&=&\!
\f{H^2(1\!-\!\epsilon)^2}{16\pi^2}\Big(\nu^2\!-\!\frac14\Big)
\Bigg[\!-\!\f{2\mu^{D-4}}{D\!-\!4} -c_\nu
+\ln\bigg(\f{4\pi\mu^2}{(1\!-\!\epsilon)^2H^2}
            \Big(1\!-\!\frac{1}{\zeta}\Big)^2\bigg)
  \!+\! 2\zeta - \zeta^2
\nonumber
\\
&&\hskip 3.5cm
 -\ln\Big(1-\frac{1}{\zeta^2}\Big)
-\f{\zeta^{1-2\nu}}{\nu\!-\!\frac12}
\times{}_2F_1\Big(1,\nu\!-\!\f12;\nu\!+\!\f12;\f{1}{\zeta^2}\Big)
\nonumber
\\
&&\hskip 3.5cm
-\ln\big(1-\zeta^2\big)
-\f{\zeta^{3+2\nu}}{\nu\!+\!\frac32}
\times{}_2F_1\Big(1,\nu\!+\!\f32;\nu\!+\!\f52;\zeta^2\Big)
\Bigg]
\,.\qquad
\label{prop:full:resum}
\end{eqnarray}
We would like to interpret this result as the analytic extension
of Eq.~(\ref{prop:full}) to arbitrary (complex) $\nu$. In order to
do this uniquely we need to specify the Riemann sheet of both the
logarithm and of the hypergeometric function in the second line
for $|\zeta|\leq 1$ and in the third line for  $|\zeta|\geq 1$
(recall that the logarithm has a branch cut along the negative
argument and that the hypergeometric function has a branch cut
running along positive arguments $z\geq 1$). It turns out that the
following cut prescription uniquely specifies the analytic
extension to arbitrary $\nu$:
\begin{eqnarray}
 \sum_{n=1}^{\nu-3/2}\frac{\zeta^{-2n}}{n}
     &\stackrel{|\zeta|< 1}{\longrightarrow}&
 -\f{1}{2}\sum_{\pm}\ln\Big(1-\frac{1}{\zeta^2\pm i\varepsilon}\Big)
-\f{1}{2}\sum_{\pm}\f{\zeta^{1-2\nu}}{\nu\!-\!\frac12}
\times{}_2F_1\Big(1,\nu\!-\!\f12;\nu\!+\!\f12;\f{1}{\zeta^2\pm
i\varepsilon}\Big) \nonumber
\\
     &=&
 -\ln\Big(\frac{1-\zeta^2}{\zeta^2}\Big)
-\f{\zeta^{3-2\nu}}{\frac32\!-\!\nu}
\times{}_2F_1\Big(1,\f32\!-\!\nu;\f52\!-\!\nu;\zeta^2\Big)
 + \pi\tan(\pi\nu)
\label{prop:analytic ext:1}
\\
\sum_{n=1}^{\nu+1/2}\frac{\zeta^{2n}}{n}
     &\stackrel{|\zeta|> 1}{\longrightarrow}&
-\frac12\sum_\pm\ln\Big(1-(\zeta^2\pm i\varepsilon)\Big)
-\frac12\sum_\pm\f{\zeta^{3+2\nu}}{\nu\!+\!\frac32}
\times{}_2F_1\Big(1,\nu\!+\!\f32;\nu\!+\!\f52;\zeta^2\pm i\varepsilon\Big)
\nonumber
\\
     &=&
 -\ln\Big(\frac{1-1/\zeta^2}{1/\zeta^2}\Big)
-\f{\zeta^{2\nu+1}}{-\frac12\!-\!\nu}
\times{}_2F_1\Big(1,-\f12\!-\!\nu;\f12\!-\!\nu;\frac{1}{\zeta^2}\Big)
 + \pi\tan(\pi\nu)
\,,
\label{prop:analytic ext:2}
\end{eqnarray}
where $\varepsilon>0$ is an infinitesimal parameter. Notice that
the results in~(\ref{prop:analytic ext:1}--\ref{prop:analytic
ext:2}) are independent on the cut prescription.
 We are now ready to write down a complete expression for
the propagator at coincidence. In the accelerating case
($\epsilon<1,\zeta=p/(aH)<1$) from Eqs.~(\ref{prop:full:resum})
and~(\ref{prop:analytic ext:1}) we thus have,
\begin{eqnarray}
i\Delta(x;x) \!&=&\!
\f{H^2(1\!-\!\epsilon)^2}{16\pi^2}\Big(\nu^2\!-\!\frac14\Big)
\Bigg[\!-\!\f{2\mu^{D-4}}{D\!-\!4} -c_\nu + \pi\tan(\pi\nu)
+\ln\bigg(\f{4\pi\mu^2}{(1\!-\!\epsilon)^2H^2(1\!+\!\zeta)^2}\bigg)
\nonumber
\\
&&\hskip 3.5cm
  +\,2\zeta - \zeta^2
-\sum_\pm\f{\zeta^{3\pm 2\nu}}{\frac32\!\pm\!\nu}
\times{}_2F_1\Big(1,\f32\!\pm\!\nu;\f52\!\pm\!\nu;\zeta^2\Big)
\Bigg]
\label{prop:full:resum:accel}
\,.\qquad
\end{eqnarray}
Notice the symmetry $\nu\rightarrow -\nu$ in all $\zeta$ dependent
terms (although, one should keep in mind that we assumed $\nu>0$).
This symmetry has been observed when the infrared is regulated by
placing the Universe in a comoving box~\cite{Janssen:2008px}. On
the other hand, in the decelerating case ($\epsilon>1,\zeta>1$) we
get from  Eqs.~(\ref{prop:full:resum}) and~(\ref{prop:analytic
ext:2}),
\begin{eqnarray}
i\Delta(x;x) \!&=&\!
\f{H^2(1\!-\!\epsilon)^2}{16\pi^2}\Big(\nu^2\!-\!\frac14\Big)
\Bigg[\!-\!\f{2\mu^{D-4}}{D\!-\!4} -c_\nu + \pi\tan(\pi\nu)
+\ln\bigg(\f{4\pi\mu^2}{(1\!-\!\epsilon)^2H^2(1\!+\!\zeta)^2}\bigg)
\nonumber
\\
&&\hskip 3.5cm +\,2\zeta - \zeta^2 -\sum_\pm\f{\zeta^{1\mp
2\nu}}{-\frac12\!\pm\!\nu}
\times{}_2F_1\Big(1,-\f12\!\pm\!\nu;\f12\!\pm\!\nu;\frac{1}{\zeta^2}\Big)
\Bigg] \label{prop:full:resum:decel} \,.\qquad
\end{eqnarray}
Just like the accelerating case~(\ref{prop:full:resum:accel}),
this expression also exhibits a $\nu\rightarrow -\nu$ symmetry in
all $\zeta$ dependent terms. It is worth noting that in the limit
when $\nu$ is a half integer, $\nu\rightarrow N+1/2$ ($N = 0,\pm
1,\pm 2,..$), both coincident
propagators~(\ref{prop:full:resum:accel})
and~(\ref{prop:full:resum:decel}) are finite. In the special case
$N=0$, both are simply zero, while for all other cases the simple
pole in the tangent cancels against the simple pole in the
hypergeometric function.

 From Eqs.~(\ref{prop:full:resum:accel}--\ref{prop:full:resum:decel})
we can easily extract the leading order late time $(a\rightarrow
\infty$) behavior of the coincident propagator in the accelerating
case ($a H\rightarrow \infty$, such that $\zeta\rightarrow 0$) and
decelerating case ($a H\rightarrow 0$, such that $\zeta\rightarrow
\infty$):
\begin{equation}
 i\Delta(x;x) \propto H^2 \zeta^{3-2\nu}
   \quad \big(\epsilon<1,\nu>\f32\,\big)
\, ;\qquad
 i\Delta(x;x) \propto H^2 \zeta^{1+2\nu}
\quad \big(\epsilon>1,\nu > \f12\,\big) \,.
\end{equation}
In the special case when $\nu=3/2$ the dependence on $\zeta$ in
the accelerating case is logarithmic.

\subsection{Numerical check of the analytic extension}
\begin{figure}[th]
\begin{center}
\includegraphics[width=6in]{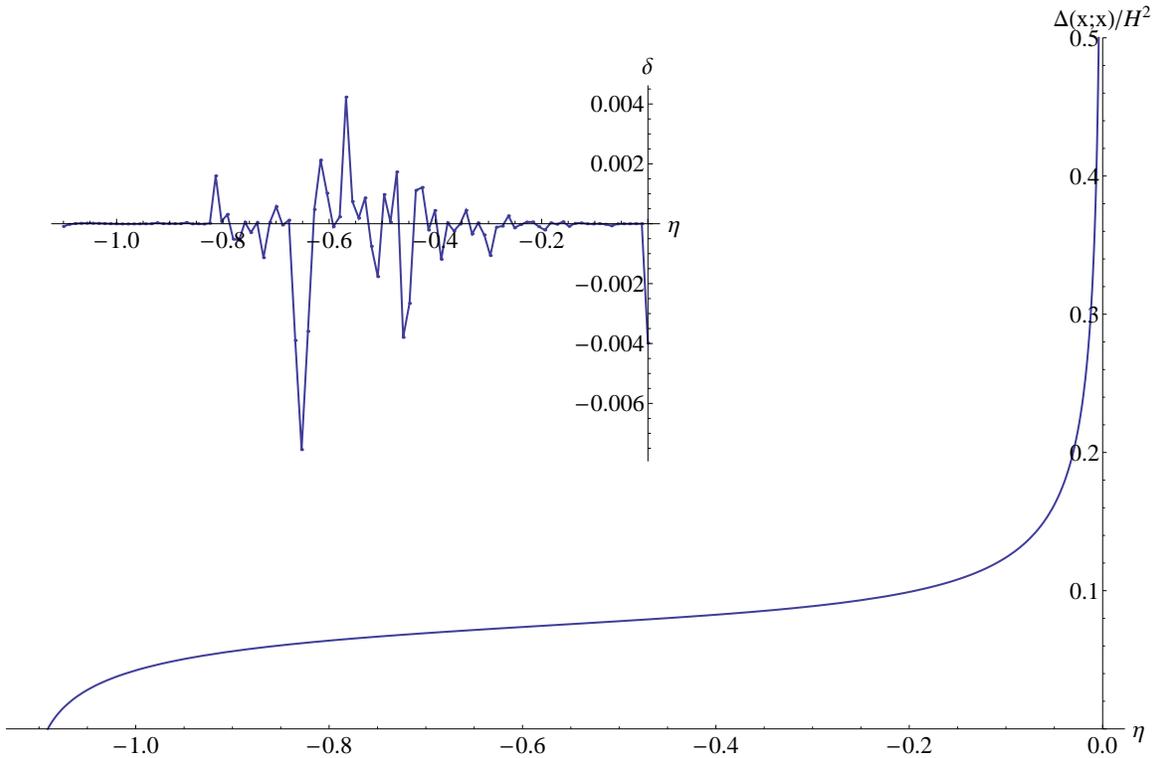}
\caption{The main plot shows the coincident propagator
(\ref{prop:full:resum:accel}), rescaled by $H^2$ versus conformal
time (\ref{conformal}). The inset shows the difference between the
analyticially calculated result and a direct numerical integration
of (\ref{propdef}), using the mode functions and the $\alpha$ and
$\beta$ coefficients in terms of Hankel functions
(\ref{mode_sol}). The parameters chosen are $p=1$, $a=4$,
$\nu=1.7$, $\epsilon=0.1$, $\Lambda=10^4$, $k_0=10^{-6}$. We find
for this case that $c_\nu\approx\psi(2\nu-2)+0.348$.
}\label{fig7}
\end{center}
\end{figure}

To test whether the procedure described above to extend the
propagator at half integer $\nu$ to all values of $\nu$, we have
performed some numerical checks. To get a numerical result for the
propagator, we use the mode functions in terms of the Hankel
function, as in (\ref{mode_sol}). From those we can easily get the
general expression for $\alpha$ and $\beta$ from the first lines
of (\ref{alphabeta}). When evaluated numerically Eq.~(\ref{propdef})
then gives an integral which, after the appropriate
regularization, corresponds to (\ref{prop:full:resum:accel}).
This is, of course, true up to the $1/(D-4)$ term, which is removed by
renormalization. There are two ultraviolet divergences: a
quadratic one and a logarithmic one. We can easily read off both
divergences from (\ref{UV:divergence}). We know those expressions
for the divergences are also valid for non half integer $\nu$. The
quadratic divergence we can simply subtract from our numerical
result. In order to correctly regulate the logarithmic divergence,
we choose an ultraviolet cutoff $\Lambda$ for our
numerical integrals and realize from (\ref{UV:divergence}) that we
must identify
\begin{equation}
    \ln(\Lambda)=\ln(a \mu\sqrt{\pi})+1-\f{1}{2}\gamma_E.
\end{equation}
To smoothen the numerical procedure, we also choose an infrared
cutoff $k_0$. Apart from the $1/(D-4)$ term, the numerically integrated
result should now correspond to (\ref{prop:full:resum:accel})
for all values of $\nu$, up to the undetermined constant $c_\nu$
in the analytic result~(\ref{prop:full:resum:accel}).
We fix this constant for a certain fixed, but arbitrary set of
parameters. As an example we show the result for $\nu=1.7$ in
figure \ref{fig7}. The main plot shows the behavior of the
coincident propagator divided by the Hubble scale
$\Delta(x;x)/H^2$ versus conformal time defined by
\begin{equation}\label{conformal}
    \eta=-\f{1}{(1-\epsilon)Ha}.
\end{equation}
For the matching point we choose $p=1$. Since $\epsilon=0.1$, we
thus find that $\hat{\eta}\approx -1.1$. The inset in the figure
shows the difference between the numerically calculated propagator
and the analytic solution (\ref{prop:full:resum:accel}). We thus
find that, up to a numerical noise of the order of 1$\%$ or less,
the two results agree. Such a noise level is not unreasonable given the
complexity of the numerical integrations. We have also checked
several other values of $\nu$, which give similar results.
Albeit we do not have a rigorous mathematical proof,
these numerical results strongly suggest that the analytic extension
to all $\nu$ made in
(\ref{prop:full:resum:accel}) and (\ref{prop:full:resum:decel}) is correct.

\section{Energy momentum tensor}
\label{s_stress}

The energy momentum tensor is defined by
\begin{equation}
    T_{\mu\nu}\equiv-\f{2}{\sqrt{-g}}\f{\delta S}{\delta
    g^{\mu\nu}}.
\end{equation}
By varying the action~(\ref{action}) we obtain
\begin{equation}\label{TMN_full}
    T_{\mu\nu}=
    (\partial_\mu\phi)\partial_\nu\phi
  -\f{1}{2}g_{\mu\nu}g^{\alpha\beta}(\partial_\alpha\phi)\partial_\beta\phi
    + \xi(R_{\mu\nu}-\f{1}{2}g_{\mu\nu}R)\phi^2
    - \xi(\nabla_\mu\nabla_\nu-g_{\mu\nu}\Box)\phi^2.
\end{equation}
We take the trace to obtain
\begin{equation}
    T^\mu{}_\mu =
    -\f{D-2}{2}\Big((\partial^\mu\phi)\partial_\mu\phi
       + R\xi\phi^2\Big)+(D-1)\xi\Box\phi^2
\label{Tq:trace}
\end{equation}
and, using the equation of motion (\ref{EOM}) for $\phi$, we find
that the one loop expectation value is given by
\begin{equation}
T_q\equiv \langle \Omega\big|T^\mu{}_\mu\big|\Omega\rangle
 = -\f14 \Big((D-2)-4(D-1)\xi\Big)\Box i\Delta(x;x)
\,.
\label{Tmn:trace}
\end{equation}
Since any quantum correction will respect the symmetries of the
underlying spacetime, we know that we should be able to write (we
use a subscript $q$ to indicate we are considering the quantum
corrections to the stress energy tensor)
\begin{equation}
 \langle\Omega\big| T^\mu{}_\nu\big|\Omega\rangle =
\mathrm{diag}\Big(-\rho_q,\underbrace{p_q,p_q,..,p_q}_{D-1}\Big)
\end{equation}
and then we can calculate $\rho_q$ and $p_q$ from $T_q$ using the
covariant stress energy conservation, which we write as
\begin{equation}
    \f{1}{H}\dot\rho_q + D\rho_q= \rho_q-(D-1)p_q = -T_q
\,.
\label{conservation}
\end{equation}
 Now an important question is whether the energy density in
the quantum corrections can dominate over the
energy density in the classical background. The background
Friedmann equations (\ref{Fried}) tell us the scaling of energy density
\begin{equation}
 \rho_b = \frac{\hat\rho_{b}}{a^{3(1+w_b)}}
\label{rho_b}
\end{equation}
where $w_b$ denotes the equation of state parameter of the
background and $\hat{\rho}_b=\rho_b(\hat{t}\,)$. Since the
background energy density is responsible for the expansion of the
Universe, from (\ref{epsilon:constant}) we have that
\begin{equation}
w_b \equiv\f{p_b}{\rho_b}= -1+\f{2}{3}\epsilon
 \label{w_b} \,.
\end{equation}
Furthermore, we know from the conservation
equation~(\ref{conservation}), which of course holds for any
stress energy tensor, that the energy density of the quantum
contribution scales in general with the scale factor as
\begin{equation}
    \rho_q = \f{\hat\rho_{q}}{a^{3(1+w_q)}}
\,,
\label{rho_q}
\end{equation}
where $w_q=p_q/\rho_q$ is the the equation of state parameter for
quantum matter contribution and $\hat\rho_{q}$ is the energy
density at the matching, which is typically small when compared to
the background density at the matching, $\hat\rho_{q}\sim \hat H^4
\sim (\hat H/M_P)^2 \hat\rho_{b}$. Here $M_P=(8\pi
G_N)^{-1/2}\simeq 2.4\times 10^{18}~{\rm GeV}$ is the reduced
Planck mass, $\hat H=H(\hat t\,)$ is the Hubble parameter at the
matching. For simplicity we took the scale factor at the matching
to be equal to unity, $\hat a = a(\hat t\,) =1$.

 From Eqs.~(\ref{rho_b}--\ref{rho_q}) we see that
\begin{equation}
  \frac{\rho_q}{\rho_b} = \frac{\hat\rho_{q}}{\hat\rho_{b}}a^{3(w_b-w_q)}
                       \sim \Big(\f{\hat H}{M_p}\Big)^2 a^{3(w_b-w_q)}
\,,
\label{rho_q-rho_b}
\end{equation}
which implies that the quantum contribution to the stress energy
grows with respect to the background contribution whenever
\begin{equation}
  w_q < w_b
\,.
\label{w_q-w_b}
\end{equation}
Hence the principal task of our investigation is to find out
whether this condition is ever met. If the answer is {\it
affirmative} then -- in the light of Eq.~(\ref{rho_q-rho_b}) -- we
would be lead to the conclusion that (after a sufficiently long
time) the energy of quantum fluctuations would eventually dominate
the energy of the background, leading to a strong quantum
backreaction. Such a strong backreaction might then significantly
change the evolution of the Universe. Of course this would then in
principle ruin our {\it Ansatz} that $\epsilon$ is constant, and
thus one should be careful in interpreting the result. If the
backreaction is large, the critical time $t_c$ after which quantum
contributions start to dominate can be estimated to be,
\begin{equation}
 t_c = \hat t \Big(\f{\hat \rho_b}{\hat \rho_q}\Big)^\f{1+w_b}{2(w_b-w_q)}
     \sim \hat t \Big(\f{M_P}{\hat H}\Big)^\f{1+w_b}{w_b-w_q}
\qquad (w_q<w_b)
\,.
\label{tc}
\end{equation}
If however $w_q>w_b$, the energy density due to the quantum
contributions will be negligible at all times. But before we
address this question, we must renormalize the stress-energy,
which is what we do next.

\section{Renormalisation}\label{reormalization}

 The only divergence arising in the one-loop stress energy tensor
is induced by the divergent part of the coincident
propagator~(\ref{prop:full:resum}), which we write in the form
\begin{equation}
  i\Delta(x;x)_{\rm div} =
-\f{H^2[(D\!-\!2)-4(D\!-\!1)\xi](D\!-\!2\epsilon)}
       {32\pi^2}\f{\mu^{D-4}}{D\!-\!4}
\,,
\label{prop:coincident div}
\end{equation}
where we took account of Eq.~(\ref{nu}). Now from the trace
equation~(\ref{Tmn:trace}) we get for the divergent contribution
to the stress energy trace,
\begin{eqnarray}
 (T_q)_{\rm div} &=& \f{[(D\!-\!2)-4(D\!-\!1)\xi]^2(D\!-\!2\epsilon)
          (D\!-\!1\!-3\epsilon)\epsilon}{64\pi^2}
                \f{\mu^{D-4}H^4}{D\!-\!4}
\label{Tq:div}
\\
 &&\hskip -1cm
=\, \f{3(1\!-\!6\xi)^2(2\!-\!\epsilon)(1\!-\!\epsilon)\epsilon}{8\pi^2}
                \f{\mu^{D-4}H^4}{D\!-\!4}
     + \f{(1\!-\!6\xi)\epsilon  H^4}{16\pi^2}
      \Big[(19\!-\!23\epsilon\!+\!6\epsilon^2)
          - 6\xi(15\!-\!17\epsilon\!+\!4\epsilon^2)\Big]
\,,
\nonumber
\end{eqnarray}
where we took account of
$\Box H^2 = -(\partial_t + (D-1)H)\partial_t H^2
          = 2\epsilon(D-1-3\epsilon)H^4$.

  It is known that this theory can be renormalized using only one
  counterterm, proportional to $R^2$. Indeed, taking a functional derivative
with respect to $g^{\mu\nu}$ of the $R^2$ counterterm action results
in~\cite{Janssen:2008px}
\begin{equation}
({\tt ct})_{\mu\nu}
\equiv  -\frac{2}{\sqrt{-g}}\frac{\delta}{\delta g^{\mu\nu}}\int d^D x\sqrt{-g}
         \alpha R^2
= \alpha (4\nabla_\mu\nabla_\nu R - 4g_{\mu\nu}\square R
          + g_{\mu\nu} R^2 -4RR_{\mu\nu})
\,.
\end{equation}
Multiplying by $g^{\mu\nu}$ results in the trace
\begin{eqnarray}
({\tt ct})_\mu^{\;\mu}
&=& \alpha \Big(-4(D\!-\!1)\square R + (D\!-\!4)R^2\Big)
\nonumber\\
&=& \alpha (D\!-\!1)^2(D\!-\!2\epsilon)H^4
\Big(-8\epsilon(D\!-\!1-3\epsilon) + (D\!-\!4)(D\!-\!2\epsilon)\Big)
\nonumber\\
&=& - 432\alpha(2\!-\!\epsilon)(1\!-\!\epsilon)\epsilon H^4
  + 36\alpha (4\!-\!34\epsilon\!+\!35\epsilon^2\!-\!8\epsilon^3)(D\!-\!4) H^4
\nonumber\\
&+&{\cal O}((D\!-\!4)^2)
\,.
\label{ct:trace}
\end{eqnarray}
Upon comparing this with Eq.~(\ref{Tq:div}) we can read off $\alpha$ that
renormalises the theory,
\begin{equation}
  \alpha = \f{(1-6\xi)^2}{1152\pi^2}\f{\mu^{D-4}}{D\!-\!4} + \alpha_f
\,,
\end{equation}
where $\alpha_f$ is an arbitrary but finite constant. One can
easily show that this choice of $\alpha$ renders all components of
$(T_\mu^{\;\nu})_q$ finite. Combining the two
contributions~(\ref{Tq:div}) and~(\ref{ct:trace}), we get the
following finite expression coming from the UV divergent
contribution, which we indicate with a superscript (1)
\begin{eqnarray}
 T_q^{(1)} &=&  (T_q)_{\rm div} + ({\tt ct})_\mu^{\;\mu}
\nonumber\\
    &=& \f{(1\!-\!6\xi)(2\!-\!\epsilon)^2H^4}{32\pi^2}
              \Big((1\!-\!6\xi)+4\f{\epsilon(1-\epsilon)}{2-\epsilon}\Big)
    - 432\alpha_f(2\!-\!\epsilon)(1\!-\!\epsilon)\epsilon H^4
\,.
\label{Tq:finite1}
\end{eqnarray}

In the following section we shall consider the contribution to the
trace of the stress energy tensor coming
from~(\ref{prop:full:resum}) to study the late time behavior of
the quantum fluid. This contribution we shall indicate with a
superscript (2).

\section{The equation of state of the quantum fluid}
\label{The equation of state of the quantum fluid}

 In this section we calculate both the pressure and energy density of
the quantum fluid that results from one loop scalar fluctuations
in our model. The general procedure is quite straightforward
(albeit somewhat tedious): we first need to act with the
d'Alembertian on the ultraviolet parts of the coincident
propagator given in
Eqs.~(\ref{prop:full:resum:accel}--\ref{prop:full:resum:decel}) to
get the trace of the stress energy tensor $T_q^{(2)}$ as defined
in Eq.~(\ref{Tmn:trace}). To get the full $T_q$ one has to add the
finite contributrion $T_q^{(1)}$ in~(\ref{Tq:finite1}) that
remained after renormalisation. When we have obtained the full
stress energy tensor, we can solve the conservation
equation~(\ref{conservation}) to obtain the quantum energy density
$\rho_q$ and quantum pressure $p_q$, from which we obtain the
quantum equation of state parameter $w_q=p_q/\rho_q$. To
facilitate this procedure we give a list of useful formulae in
Appendices A and B.

 To illustrate the procedure, we shall first calculate the
 contribution to $\rho_q$ and $p_q$ coming from the finite remainder from ultraviolet
 fluctuations~(\ref{Tq:finite1}). We denote these contributions
 by  $\rho_q^{(1)}$ and $p_q^{(1)}$.

 To calculate
$\rho_q^{(1)}$ we need to integrate Eq.~(\ref{conservation}),
which we can do with the help of Eq.~(\ref{AppB:2}) in Appendix~B.
The result is,
\begin{equation}
 \rho_q^{(1)} = - \f{(1\!-\!6\xi)(2\!-\!\epsilon)^2H^4}
                    {128\pi^2(1\!-\!\epsilon)}
              \Big((1\!-\!6\xi)+4\epsilon\Big)
    + 108\alpha_f(2\!-\!\epsilon)\epsilon H^4
\label{rho_q1}
\end{equation}
The contribution to the pressure is then simply,
\begin{eqnarray}
 p_q^{(1)} &=& \frac{T_q^{(1)}+\rho_q^{(1)}}{3}
           =  \bigg(-\f{(1\!-\!6\xi)(2\!-\!\epsilon)^2H^4}
                    {128\pi^2(1\!-\!\epsilon)}
              \Big((1\!-\!6\xi)+4\epsilon\Big)
    + 108\alpha_f(2\!-\!\epsilon)\epsilon H^4
             \bigg)\Big(-1+\frac{4}{3}\epsilon\Big)
\nonumber\\
 &=& \rho_q^{(1)}\Big(-1+\frac{4}{3}\epsilon\Big)
\label{p_q1}
\end{eqnarray}
 Of course, the one loop infrared contribution needs to be taken into
account for a complete answer.
  We shall now consider the general infrared
contributions to the quantum energy density and pressure.
We first consider the accelerated and then the decelerated case.

\subsection{Matching onto acceleration ($\epsilon<1$) for a general $\nu$}
\label{Matching onto acceleration: general}

In this section we consider in some detail the general
accelerating case ($\epsilon<1$, $\zeta<1$). The decelerating case
we shall consider in section \ref{Matching onto deceleration:
general}. The contribution to the trace of the stress-energy
tensor coming from the UV divergent terms has been calculated in
(\ref{Tq:finite1}). Our starting point for the remaining terms is
the coincident propagator~(\ref{prop:full:resum:accel}). Now
making use of Eq.~(\ref{box identity 1}) in Appendix~A the
contribution to the trace of the quantum stress-energy
tensor~(\ref{Tmn:trace}) from the remaining terms becomes:
\begin{eqnarray}
T_q^{(2)} &=&
  -\f{(1\!-\!6\xi)^2(1\!-\!\epsilon)(2\!-\!\epsilon)H^4}{16\pi^2}
\Bigg\{
   6\epsilon\Bigg[
          \frac12 \ln\bigg(\f{4\pi\mu^2}{(1\!-\!\epsilon)^2H^2(1\!+\!\zeta)^2}
                  \bigg)
                 -\f{c_\nu}{2}
 \nonumber\\
  &&\hskip 3.5cm
  -\sum_\pm \f{\zeta^{3\pm 2\nu}}{3\pm 2\nu}
                 \times{}_2F_1\Big(1,\f{3}{2}\pm\nu;\f{5}{2}\pm\nu;\zeta^2\Big)
+\f{\pi}{2}\tan(\pi\nu)
            \Bigg]
\nonumber\\
  &&\hskip 3.cm
  -\,\f{3\!-\!5\epsilon}{1\!-\!\epsilon} + \f{2(2\!-\!3\epsilon)}{1\!+\!\zeta}
  - \f{1\!-\!\epsilon}{(1\!+\!\zeta)^2}
  + 2(1\!+\!\epsilon)\zeta - \zeta^2
\label{Tq2:exact}
\\
  &&\hskip 2.5cm
   -\,\sum_\pm \Big[(3\!-\!5\epsilon)-(1\!-\!\epsilon)(3\pm 2\nu)\Big]
                \f{\zeta^{3\pm 2\nu}}{1\!-\!\zeta^2}
     +2(1\!-\!\epsilon)\sum_\pm\f{\zeta^{5\pm 2\nu}}{(1\!-\!\zeta^2)^2}
 \Bigg\}
\,.
\nonumber
\end{eqnarray}
Recall that by assumption $\nu>0$. The $\nu=1/2$ pole of
$\tan(\pi\nu)$ is harmless since in this case $\xi=1/6$, such that
the whole expression vanishes as the result of a vanishing
prefactor. In Eq.~(\ref{Tq2:exact}) we made use the derivative of
the hypergeometric function,
\begin{equation}
\zeta\f{d}{d\zeta}\f{\zeta^{2a}}{2a}\times{}_2F_1(1,a;a+1;\zeta^2)
  =\f{\zeta^{2a}}{1-\zeta^2}
\,.
\nonumber
\end{equation}
 The total stress energy trace is then obtained by adding
Eqs.~(\ref{Tq:finite1}) and~(\ref{Tq2:exact}).\\
 A careful look at Eqs.~(\ref{Tq2:exact}) reveals a quadratic divergence
in $T_q$ at the matching time $t=\hat t$, at which
$\zeta\rightarrow 1$. This divergence is also present in the
energy density and pressure calculated below in
Eqs.~(\ref{rhoq2:exact}--\ref{pq2:exact}), and shows up as the
logarithmic divergence in the coincident
propagator~(\ref{prop:full:resum}). This divergence occurs because
of the sudden matching procedure assumed in this work, where
$\epsilon$ jumps suddenly from $\epsilon=2$ to an arbitrary
$\epsilon$, which also implies the sudden change in the Ricci
scalar and the scalar field mass parameter. Of course, in any
physically realistic situation any of those parameters will change
smoothly.
 These sudden changes induce a change in the quantum contribution
by an infinite amount at the matching. An attempt to renormalise
it away by including it into the $1/(D-4)$ subtraction in
Eq.~(\ref{UV:divergence}) would result in divergent contributions
away from the matching, which would make matters only worse. But
let us try to understand how bad the matching divergence actually
is. From Eq.~(\ref{Tq2:exact}) we see that its contribution to
$T_q$ is of the order $\sim H^4/[1-\hat{a}\hat{H}/(aH)]^2$. By
taking $t-\hat t=\delta t$ small, and requring that $\rho_q<\rho_b
=3M_P^2H^2$, where $M_P=(8\pi G)^{-1/2}$ is the reduced Planck
mass, we get that $\rho_q<\rho_b$ implies $\delta t
>1/[|1-\epsilon|M_P]$, which is of the order the Planck time.
Based on this estimate we expect that the matching divergence will
be regulated whenever a sudden matching is replaced by a smooth
matching, in which $\epsilon$ at the matching changes over a
period of time that is much longer than the Planck time. \\
 Before we calculate the quantum energy density,
we take one more look at the form of Eq.~(\ref{Tq2:exact}).
All terms containing $\nu$ dependent powers of $\zeta$
in~(\ref{Tq2:exact}) can be written as the following sum,
\begin{equation}
 (T_q^{(2)})_\nu =  \f{(1\!-\!6\xi)^2(1\!-\!\epsilon)(2\!-\!\epsilon)H^4}{16\pi^2}\sum_\pm\sum_{n=0}^\infty
  \bigg[\f{6\epsilon}{3\!\pm\!2\nu\!+\!2n}+(3\!-\!5\epsilon)
       -(1\!-\!\epsilon)(3\!\pm\!2\nu\!+\!2n)\bigg]\zeta^{3\pm2\nu+2n}
\label{Tq2:exact:2}
\end{equation}
 The contribution of this sum to the energy density
can be, up to a  $\nu$ dependent integration constant ${\rm
cte}(\nu)\zeta^4H^4$, determined by integrating
Eq.~(\ref{conservation}), which can be performed by making use
of Eq.~(\ref{AppB:6}). The result is
\begin{eqnarray}
&&\hskip -0.8cm   (\rho_q^{(2)})_\nu =
\f{(1\!-\!6\xi)^2(1\!-\!\epsilon)(2\!-\!\epsilon)H^4}{16\pi^2}\sum_\pm\sum_{n=0}^\infty
  \bigg[\f{6\epsilon}{3\!\pm\!2\nu\!+\!2n}+(3\!-\!5\epsilon)
       -(1\!-\!\epsilon)(3\!\pm\!2\nu\!+\!2n)\bigg]
\nonumber\\
     && \hskip 7cm \times     \f{\zeta^{3\pm2\nu+2n}}
              {(1\!-\!\epsilon)(\!-\!1\!\pm\!2\nu\!+\!2n)} +{\rm
              cte}(\nu)\zeta^4H^4
\label{Tq2:exact:3}
\\
&&\hskip -.5cm =\, \f{(1\!-\!6\xi)^2(2\!-\!\epsilon)H^4}{64\pi^2}
\Bigg\{\!-6\epsilon\sum_\pm \f{\zeta^{3\pm2\nu}}{3\!\pm\!2\nu}
     \times{}_2F_1\Big(1,\f32\pm\nu;\f52\pm\nu;\zeta^2\Big)
   -\,4(1\!-\!\epsilon)\sum_\pm
    \f{\zeta^{3\pm2\nu}}{1\!-\!\zeta^2}
\nonumber\\
&&\hskip 1.cm -\, 2(2\!-\!\epsilon)
     \sum_\pm \f{\zeta^{3\pm2\nu}}{\!-\!1\!\pm\!2\nu}
     \times{}_2F_1\Big(1,-\f12\pm\nu;\f12\pm\nu;\zeta^2\Big)
         +\Big((2\!-\!\epsilon)\pi\tan(\pi\nu)+d_\nu\Big)\zeta^4
\Bigg\}
\,,\;
\nonumber
\end{eqnarray}
where in the last step we chose the integration constant to be
\begin{equation}
    \rm{cte}(\nu)=\f{(1\!-\!6\xi)^2(2\!-\!\epsilon)}{64\pi^2}\Big((2\!-\!\epsilon)\pi\tan(\pi\nu)+d_\nu\Big),
\label{cte}
\end{equation}
where the tangent contribution is fixed by requiring that $\rho_q^{(2)}$ be
finite for all $\nu$, and
$d_\nu$ a $\nu$ dependent function, which is finite for all $\nu$.
Notice that the divergences at half integer $\nu$ coming
from the first hypergeometric function are cancelled by the
contribution from the tangent in (\ref{Tq2:exact}).
The integration constant~(\ref{cte}) scales as radiation, $\rho_q\propto
1/a^4$, and it can be uniquely fixed by calculating the full stress energy
tensor instead of just the trace. But in order to do that, we need
the propagator off coincidence, which we do not have at this moment.
\\
 With this we are now ready to calculate the quantum energy density and
pressure, which are obtained by integrating
Eq.~(\ref{conservation}) with the help of Eqs.~(\ref{AppB:4}) in
Appendix~B. The result for the energy density is
\begin{eqnarray}
\rho_q^{(2)} &=& \f{(1\!-\!6\xi)^2(2\!-\!\epsilon)H^4}{64\pi^2}
\Bigg\{
   6\epsilon\Bigg[
          \frac12 \ln\bigg(\f{4\pi\mu^2}{(1\!-\!\epsilon)^2H^2(1\!+\!\zeta)^2}
                  \bigg)
                 -\f{c_\nu}{2}
     - \f{6\!-\!7\epsilon}{12(1\!-\!\epsilon)}
 \nonumber\\
  &&\hskip 3.99cm
  -\sum_\pm \f{\zeta^{3\pm 2\nu}}{3\!\pm\! 2\nu}
           \!\times\!{}_2F_1\Big(1,\f{3}{2}\pm\nu;\f{5}{2}\pm\nu;\zeta^2\Big)
  +\f{\pi}{2}\tan(\pi\nu)
\nonumber\\
  &&\hskip 3.99cm
  +\, \zeta -\frac12\zeta^2 +\frac13\zeta^3
            \Bigg]
 - 2(2\!-\!\epsilon)\zeta^4\ln\Big(\f{1+\zeta}{\zeta}\Big)
 + 4(1\!-\!\epsilon)\f{\zeta^4}{1+\zeta}
\nonumber\\
&&\hskip 3.2cm
 -\, 2(2\!-\!\epsilon)
     \sum_\pm \f{\zeta^{3\pm2\nu}}{\!-\!1\!\pm\!2\nu}
     \times{}_2F_1\Big(1,-\f12\pm\nu;\f12\pm\nu;\zeta^2\Big)
\nonumber\\
&&\hskip 3.2cm
         +\Big((2\!-\!\epsilon)\pi\tan(\pi\nu)+d_\nu\Big)\zeta^4
      -\,4(1\!-\!\epsilon)\sum_\pm
        \f{\zeta^{3\pm2\nu}}{1\!-\!\zeta^2}
\Bigg\}
\,,
\label{rhoq2:exact}
\end{eqnarray}
with $\nu>0$.
The total energy density is obtained by adding this to the ultraviolet
contribution~(\ref{rho_q1}). The quantum pressure
$p_q$ is then simply a sum of
\begin{equation}
  p_q^{(2)} = \frac13(T_q^{(2)}+\rho_q^{(2)})
\label{pq2:exact}
\end{equation}
and the ultraviolet contribution $p_q^{(1)}$ in Eq.~(\ref{p_q1}).
The equation of state parameter $w_q$ is ,
\begin{equation}
  w_q = \f{p_q}{\rho_q}.
\label{wq}
\end{equation}

 In other to find out whether the quantum contribution can become important,
we need to investigate whether the criterion~(\ref{w_q-w_b}) for
the growth of quantum contribution with respect to the classical
contibution is ever met (see also Eq.~(\ref{tc}) above). Rather
then studying $w_q(t)$ in its full generality, we shall consider
only the leading (late time) behaviour of $w_q$. For this we need
the leading order behavior of  $T_q$,  $\rho_q$ and $p_q$.
Remember that in this section we consider an accelerating universe
such that at late times $\zeta\rightarrow 0$.

 \subsubsection{The case when $\epsilon<1$, $0<\nu<3/2$}
 \label{The case when epsilon<1, nu<3/2}

 From Eqs.~(\ref{Tq:finite1}), (\ref{Tq2:exact}),
(\ref{rho_q1}) (\ref{p_q1}), (\ref{rhoq2:exact})  and~(\ref{pq2:exact})
it follows that when $0<\nu<3/2$ the leading order late time contribution
to the one loop stress energy trace, energy density and pressure are,
\begin{eqnarray}
  T_q &\stackrel{\zeta\rightarrow 0}{\longrightarrow}&
  -\f{(1\!-\!6\xi)^2(1\!-\!\epsilon)(2\!-\!\epsilon)H^4}{16\pi^2}
\bigg\{6\epsilon\bigg[ \ln\Big(\f{H_0}{H}\Big)
                  + \f{\epsilon}{4(1\!-\!\epsilon)}
                \bigg]
\bigg\}
\label{acc:Tq}
\\
  \rho_q &\stackrel{\zeta\rightarrow 0}{\longrightarrow}&
  \f{(1\!-\!6\xi)^2(2\!-\!\epsilon)H^4}{64\pi^2}
\bigg\{6\epsilon\ln\Big(\f{H_0}{H}\Big)
\bigg\}
\label{acc:rhoq}
\\
  p_q &\stackrel{\zeta\rightarrow 0}{\longrightarrow}&
  \f{(1\!-\!6\xi)^2(2\!-\!\epsilon)H^4}{64\pi^2}
\bigg\{6\epsilon\bigg[\Big(-1+\f43\epsilon\Big) \ln\Big(\f{H_0}{H}\Big)
                     -\f{\epsilon}{3}
               \bigg]
\bigg\}
\,,
\label{acc:pq}
\end{eqnarray}
where
\begin{eqnarray}
  \ln(H_0) &=& \f12\ln\bigg(\f{4\pi\mu^2}{(1\!-\!\epsilon)^2}\bigg)
             -\f{c_\nu}{2}
             - \f{3\!-\!5\epsilon}{6(1\!-\!\epsilon)}
             - \f{\epsilon}{4(1\!-\!\epsilon)}
             +\f\pi{2}\tan(\pi\nu)
\nonumber\\
           &&  -\,\f{2\!-\!\epsilon}{3(1\!-\!\epsilon)}
                       \Big(\f{1}{4\epsilon}+\f{1-\epsilon}{(2-\epsilon)(1\!-\!6\xi)}\Big)
            +\f{1152\pi^2}{(1-6\xi)^2}\alpha_f
\,.
\label{acc:H0}
\end{eqnarray}
These relations then imply for the equation of state parameter
$w_q=p_q/\rho_q$:
\begin{equation}
  w_q \stackrel{\zeta\rightarrow 0}{\longrightarrow}
    \Big(-1+\f43\epsilon\Big)
     - \f{\epsilon}{3\ln\big(H_0/H\big)}
     \rightarrow -1+\f43\epsilon
\,.
\label{acc:wq}
\end{equation}
The last implication follows from the observation
that at late times $a\rightarrow \infty$,
$H\propto a^{-\epsilon}\rightarrow 0$,
such that formally $\ln(H_0/H)\rightarrow \infty$ as long as $\epsilon>0$.
When $\epsilon=0$ one recovers the well known result, $w_q=-1$.
The results~(\ref{acc:Tq}--\ref{acc:wq}) also apply when $\nu$ is
imaginary, or more generally whenever $\Re[\nu]<3/2$.

 \subsubsection{The case when $\epsilon<1$, $\nu>3/2$}
 \label{The case when epsilon<1, nu>3/2}

 When $\nu>3/2$ in general the lowest $\nu$ dependent power
in~(\ref{Tq2:exact}) and~(\ref{rhoq2:exact}) dominates the
stress energy tensor at late times. From the more convenient
form~(\ref{Tq2:exact:2}--\ref{Tq2:exact:3}) expressed as series
we can read off the dominant contribution to $T_q$, $\rho_q$ and $p_q$:
\begin{eqnarray}
  T_q &\stackrel{\zeta\rightarrow 0}{\longrightarrow}&
  \f{(1\!-\!6\xi)^2(1\!-\!\epsilon)(2\!-\!\epsilon)H^4}{16\pi^2}
\bigg[\f{6\epsilon}{3\!-\!2\nu}+(3\!-\!5\epsilon)
       -(1\!-\!\epsilon)(3\!-\!2\nu)\bigg]\zeta^{3-2\nu}
\,,
\label{acc:Tq:2}
\\
  \rho_q &\stackrel{\zeta\rightarrow 0}{\longrightarrow}&
  -\f{(1\!-\!6\xi)^2(2\!-\!\epsilon)H^4}{64\pi^2}
\bigg[\f{6\epsilon}{3\!-\!2\nu}+(3\!-\!5\epsilon)
       -(1\!-\!\epsilon)(3\!-\!2\nu)\bigg]\f{4\zeta^{3-2\nu}}{1\!+\!2\nu}
 \label{acc:rhoq:2}
\\
  p_q &\stackrel{\zeta\rightarrow 0}{\longrightarrow}&
  -\f{(1\!-\!6\xi)^2(2\!-\!\epsilon)H^4}{64\pi^2}
\bigg[\f{6\epsilon}{3\!-\!2\nu}+(3\!-\!5\epsilon)
       -(1\!-\!\epsilon)(3\!-\!2\nu)\bigg]
            \f{4[\epsilon-2(1\!-\!\epsilon)\nu]}{3[1\!+\!2\nu]}\zeta^{3-2\nu}
.
\label{acc:pq:2}
\end{eqnarray}
From this we conclude,
\begin{equation}\label{acc:wq:2}
\begin{split}
 w_q &\stackrel{\zeta\rightarrow 0}{\longrightarrow}
   -\f{2}{3}(1\!-\!\epsilon)\nu + \f{\epsilon}{3}
\qquad(3/2<\nu)\\
&\stackrel{\zeta\rightarrow 0}{\longrightarrow}
   w_b+\f{1}{3}\Big((3-\epsilon)-\sqrt{(3-\epsilon)^2-24(2-\epsilon)\xi}\Big) \,.
\end{split}
\end{equation}
from which we see that $w_q<w_b$ if $\xi$ is negative and
$w_q>w_b$ is $\xi$ is positive, for all relevant $\epsilon$
(remember that this analysis holds for $\epsilon<1$). Thus we find
that the quantum contribution to the energy density grows with
respect to the backgound energy density for $\xi<0$.\\

This result is correct however, provided the leading order
contributions~(\ref{acc:Tq:2}--\ref{acc:pq:2}) do not vanish,
which is the case when the expression in the square brackets does not
vanish, {\it i.e.} when
\begin{equation}
\f{6\epsilon}{3\!-\!2\nu}+(3\!-\!5\epsilon)
       -(1\!-\!\epsilon)(3\!-\!2\nu)  \neq 0
\,.
\label{acc:special:nu}
\end{equation}
We shall now study in some detail the special case when
(\ref{acc:special:nu}) vanishes.

\subsubsection{The special case when $\epsilon<1$, $\xi=0$}
 \label{The special case when epsilon<1, xi=0}

 It is instructive to observe that~(\ref{acc:special:nu}) vanishes for the
following two values of $\nu$:
\begin{equation}
 \nu \in \left\{0\;,\; \f{3\!-\!\epsilon}{2(1\!-\!\epsilon)}\right\}
\,,
\label{acc:special:nu:2}
\end{equation}
or equivalently when
\begin{equation}
 \xi \in \left\{\f{(3\!-\!\epsilon)^2}{24(2\!-\!\epsilon)}\;,\; 0\right\}
\,.
\end{equation}
The first value in~(\ref{acc:special:nu:2}) is irrelevant since
we consider $\nu>3/2$. The second value however might be interesting. It
corresponds to the minimal coupling of the scalar field to
curvature, $\xi=0$. If $3/2<\nu<5/2$ (or equivalently
$0<\epsilon<1/2)$, the leading order contribution in this case is
logarithmic and the above analysis presented in
subsection~\ref{The case when epsilon<1, nu<3/2} applies and we
have ({\it cf.} Eq.~(\ref{acc:wq})),
\begin{equation}
  w_q  \stackrel{\zeta\rightarrow 0}{\longrightarrow} -1+\f43\epsilon
\qquad (\xi=0\,,\,3/2<\nu<5/2\,,\,0<\epsilon<1/2)
\,.
\label{acc:wq:special:nu:1}
\end{equation}
When, on the other hand, $\nu>5/2$ ($1/2<\epsilon<1$), the
dominant contribution comes from the subdominant term in the $\nu$
dependent series~(\ref{Tq2:exact:2}--\ref{Tq2:exact:3}). Since we
study the case where $\xi=0$ we have from (\ref{nu}) that
$\nu=(3-\epsilon)/[2(1-\epsilon)]$ and we can write the stress
energy contributions as,
\begin{eqnarray}
  T_q &\stackrel{\zeta\rightarrow 0}{\longrightarrow}&
  \f{(1\!-\!6\xi)^2(1\!-\!\epsilon)^2(2\!-\!\epsilon)(1\!+\!\epsilon)H^4}
    {16\pi^2(1\!-\!2\epsilon)}
\zeta^{5-2\nu} \,, \label{acc:Tq:special:nu}
\\
  \rho_q &\stackrel{\zeta\rightarrow 0}{\longrightarrow}&
   -\f{(1\!-\!6\xi)^2(1\!-\!\epsilon)^2(2\!-\!\epsilon)(1\!+\!\epsilon)H^4}
    {32\pi^2(1\!-\!2\epsilon)}
\zeta^{5-2\nu} \quad (5/2<\nu,1/2<\epsilon<1)\qquad
\label{acc:rhoq:special:nu}
\\
  p_q &\stackrel{\zeta\rightarrow 0}{\longrightarrow}&
  \f{(1\!-\!6\xi)^2(1\!-\!\epsilon)^2(2\!-\!\epsilon)(1\!+\!\epsilon)H^4}
    {96\pi^2(1\!-\!2\epsilon)}
\zeta^{5-2\nu} \;. \label{acc:pq:special:nu}
\end{eqnarray}
Thus the equation of state parameter is in this case,
\begin{equation}
  w_q  \stackrel{\zeta\rightarrow 0}{\longrightarrow} - \f13
\qquad (\xi=0\,,\,\nu>5/2\,,\,1/2<\epsilon<1)
\,.
\label{acc:wq:special:nu:2}
\end{equation}

In figures \ref{fig1}, \ref{fig2} and \ref{fig3} we illustrate the
results of this section (these figures also already show the
results for the decelerating case, $\epsilon>1$ which we shall
discuss in section \ref{Matching onto deceleration: general}).
From the previous discussion we found that in an accelerating
universe  quantum effects can become important if $\xi<0$ while
they will always be subdominant for $\xi>0$. In figure~\ref{fig1}
we show the borderline behavior: $\xi=0$. The equation of state
parameter of quantum fluid $w_q = -1+4\epsilon/3$ (when
$0<\epsilon<1/2$, conform (\ref{acc:wq:special:nu:1})) and $w_q =
-1/3$ (when $1/2<\epsilon<1$, conform (\ref{acc:wq:special:nu:2}))
is always larger than the corresponding background parameter $w_b
= -1 + 2\epsilon/3$, implying that the criterion~(\ref{w_q-w_b})
is never met and the quantum contribution can never become
important. If, on the other hand, $\xi\neq 0$  the situation
changes drastically.  In figure~\ref{fig2} we show $w_q$ for
positive values of $\xi$. The curves are given by
$w_q=-1+\f{4}{3}\epsilon$ for $\nu<3/2$ (conform (\ref{acc:wq}))
and $w_q = -\f{2}{3}(1-\epsilon)\nu+\f{\epsilon}{3}$ for $\nu>3/2$
(conform (\ref{acc:wq:2})). We indeed see that we always have
$w_q>w_b$, such that the energy density due to the quantum effects
dilutes faster then the background energy density. In figure
\ref{fig3} we show $w_q$ for negative values of $\xi$. Since for
negative $\xi$ we always have that $\nu>3/2$ we have (conform
(\ref{acc:wq:2})) that $w_q =
-\f{2}{3}(1-\epsilon)\nu+\f{\epsilon}{3}$. We indeed see that
$w_q<w_b$ in this case and thus the energy density in the quantum
contribution will dominate at late enough times. The time it takes
for the quantum contribution to become important can be estimated
from Eq.~(\ref{tc}) to be
\begin{equation}
 t_c \sim \f{1}{(1-\epsilon)\hat H}
   \Big(\f{M_P}{\hat H}\Big)^\f{\epsilon(3-\epsilon)}{-6\xi(2-\epsilon)}
\qquad (\xi<0,0<\epsilon<1)
\,.
\label{tc:acc}
\end{equation}
For $0<-6\xi\ll 1$ this time can be very long, much longer than the age
of the Universe. In fact $-6\xi< 1$ can be
tuned to make $t_c$ of the order of the age of the Universe, in which
case one could relate the contribution of quantum
fluctuations to the dark energy of the Universe, and perhaps
even address the question {\it `Why now!'}
of the dark energy dominance.
The above analysis works however only for matching
radiation era onto accelerating Universes, in which the Universe
spends most of the time, and that does not correspond to the observed Universe.
Therefore, we also need to consider radiation matching
onto the decelerated universes, which is what we hurriedly do next.

\subsection{Matching onto deceleration for a general $\nu$}
\label{Matching onto deceleration: general}

 The exact result for $T_q^{(2)}$ in Eq.~(\ref{Tq2:exact})
is also correct for the decelerating universe.
Since in a decelerating universe at late times
$\zeta\rightarrow \infty$, it is more
suitable to express  $T_q^{(2)}$ in terms of the hypergeometric function
of $1/\zeta^2$. The result is,
\begin{eqnarray}
T_q^{(2)} &=&
  -\f{(1\!-\!6\xi)^2(1\!-\!\epsilon)(2\!-\!\epsilon)H^4}{16\pi^2}
\Bigg\{
   6\epsilon\Bigg[
          \frac12 \ln\bigg(\f{4\pi\mu^2}{(1\!-\!\epsilon)^2H^2(1\!+\!\zeta)^2}
                  \bigg)
                 -\f{c_\nu}{2}
 \nonumber\\
  &&\hskip 3.5cm
  -\sum_\pm \f{\zeta^{1\mp 2\nu}}{-1\pm 2\nu}
         \times{}_2F_1\Big(1,-\f{1}{2}\pm\nu;\f{1}{2}\pm\nu;\f{1}{\zeta^2}\Big)
+\f{\pi}{2}\tan(\pi\nu)
            \Bigg]
\nonumber\\
  &&\hskip 3.cm
  -\,\f{3\!-\!5\epsilon}{1\!-\!\epsilon} + \f{2(2\!-\!3\epsilon)}{1\!+\!\zeta}
  - \f{1\!-\!\epsilon}{(1\!+\!\zeta)^2}
  + 2(1\!+\!\epsilon)\zeta - \zeta^2
\label{Tq2:exact:dec}
\\
  &&\hskip 2.5cm
   -\,\sum_\pm \Big[(3\!-\!5\epsilon)-(1\!-\!\epsilon)(3\mp 2\nu)\Big]
                \f{\zeta^{3\mp 2\nu}}{1\!-\!\zeta^2}
     +2(1\!-\!\epsilon)\sum_\pm\f{\zeta^{5\mp 2\nu}}{(1\!-\!\zeta^2)^2}
 \Bigg\}
\,,
\nonumber
\end{eqnarray}
where we made use of the identity,
\begin{equation}
 \f{\zeta^{2\alpha}}{2\alpha}\times{}_2F_1\Big(1,\alpha;\alpha+1;\zeta^2\Big)=
 \f{\zeta^{2(\alpha-1)}}{2(1-\alpha)}
   \times{}_2F_1\Big(1,1-\alpha;2-\alpha;\f{1}{\zeta^2}\Big)
 +\f{\pi}{2}\cot(\pi \alpha)
\quad (\zeta^2>1)\,.\quad
\label{zeta-to-inverse}
\end{equation}
In the derivation of this identity,
we took the central value along the cut of the hypergeometric function
$\zeta^2\geq 1$ (this prescription removes a possible imaginary contribution).
Notice that for $\alpha=(3/2)\pm\nu$ the last bit in~(\ref{zeta-to-inverse})
does not contribute, since $\sum_\pm \cot[\pi (3/2)\pm \pi\nu] =0$.
The result~(\ref{Tq2:exact:dec}) can be also obtained from
Eq.~(\ref{Tmn:trace}) by acting with $\Box$ on
Eq.~(\ref{prop:full:resum:decel}),
representing a check of~(\ref{Tq2:exact:dec}).
The identity~(\ref{zeta-to-inverse}) can be also used to derive
$\rho_q^{(2)}$ from Eq.~(\ref{rhoq2:exact}),
\begin{eqnarray}
\rho_q^{(2)} &=& \f{(1\!-\!6\xi)^2(2\!-\!\epsilon)H^4}{64\pi^2}
\Bigg\{
   6\epsilon\Bigg[
          \frac12 \ln\bigg(\f{4\pi\mu^2}{(1\!-\!\epsilon)^2H^2(1\!+\!\zeta)^2}
                  \bigg)
                 -\f{c_\nu}{2}
     - \f{6\!-\!7\epsilon}{12(1\!-\!\epsilon)}
 \nonumber\\
  &&\hskip 3.99cm
  -\sum_\pm \f{\zeta^{1\mp 2\nu}}{-1\!\pm\! 2\nu}
     \!\times\!{}_2F_1\Big(1,-\f{1}{2}\pm\nu;\f{1}{2}\pm\nu;\f{1}{\zeta^2}\Big)
  +\f{\pi}{2}\tan(\pi\nu)
\nonumber\\
  &&\hskip 3.99cm
  +\, \zeta -\frac12\zeta^2 +\frac13\zeta^3
            \Bigg]
 - 2(2\!-\!\epsilon)\zeta^4\ln\Big(1+\f{1}{\zeta}\Big)
 + 4(1\!-\!\epsilon)\f{\zeta^4}{1+\zeta}
\nonumber\\
&&\hskip 3.2cm
 -\, 2(2\!-\!\epsilon)
     \sum_\pm \f{\zeta^{1\mp2\nu}}{3\!\pm\!2\nu}
     \times{}_2F_1\Big(1,\f32\pm\nu;\f52\pm\nu;\f{1}{\zeta^2}\Big)
\nonumber\\
&&\hskip 3.2cm
      +\,\Big((2\!-\!\epsilon)\pi\tan(\pi\nu)+d_\nu\Big)\zeta^4
      -4(1\!-\!\epsilon)\sum_\pm
        \f{\zeta^{3\mp2\nu}}{1\!-\!\zeta^2}
\Bigg\}
\,,
\label{rhoq2:exact:dec}
\end{eqnarray}
The one loop pressure $p_q^{(2)}$ is, as before, obtained by
inserting~(\ref{Tq2:exact:dec}) and~(\ref{rhoq2:exact:dec}) into
relation~(\ref{pq2:exact}). We are now ready to consider the late
time limit, $a\rightarrow \infty$ which implies for the
decelerating case that $\zeta\rightarrow \infty$, of the stress
energy tensor and the corresponding equation of state parameter in
a decelerating universe. Before we begin analysing particular
cases, let us observe the general structure of
Eqs.~(\ref{Tq2:exact:dec}) and~(\ref{rhoq2:exact:dec}). The curly
brackets in~(\ref{Tq2:exact:dec}) contain terms that grow $\propto
\zeta^4$ and terms with $\nu$-dependent powers of $\zeta$. The
$\nu$-dependent powers can be summarised as,
\begin{equation}
 (T_q^{(2)})_\nu = \f{(1\!-\!6\xi)^2(1\!-\!\epsilon)(2\!-\!\epsilon)H^4}{16\pi^2}\sum_\pm\sum_{n=0}^\infty
   \bigg[\f{6\epsilon}{-1\!\pm\!2\nu\!+\!2n}-(3\!-\!5\epsilon)
       -(1\!-\!\epsilon)(-1\!\pm\!2\nu\!+\!2n)
   \bigg]\zeta^{1\mp2\nu-2n}
\,.
\label{Tq:dec:nu-powers}
\end{equation}
Similarly, $\rho_q^{(2)}$ in~(\ref{rhoq2:exact:dec}) contains
terms that grow as $\propto \zeta^4$, and the terms that contain
$\nu$-dependent powers of $\zeta$, which can be summarized as,
\begin{equation}
 (\rho_q^{(2)})_\nu = \f{(1\!-\!6\xi)^2(2\!-\!\epsilon)H^4}{64\pi^2}\sum_\pm\sum_{n=0}^\infty
   \bigg[\f{6\epsilon}{-1\!\pm\!2\nu\!+\!2n}-(3\!-\!5\epsilon)
       -(1\!-\!\epsilon)(-1\!\pm\!2\nu\!+\!2n)
   \bigg]\f{4\zeta^{1\mp2\nu-2n}}{3\pm2\nu+2n}
\,.
\label{rhoq:dec:nu-powers}
\end{equation}
This form could be also obtained by
integrating~(\ref{Tq:dec:nu-powers}), representing a nontrivial
check of our result~(\ref{rhoq2:exact:dec}). It is now clear that
the leading order late time behavior of $\rho_q$ and $T_q$ is
dictated either by the term $\propto \zeta^4$ in $\rho_q$ or by a
term $\propto \zeta^{1+2\nu-2n}$ with $n=0,1,\ldots$ in
Eqs.~(\ref{Tq:dec:nu-powers}--\ref{rhoq:dec:nu-powers}). In both
cases the ultraviolet contributions~(\ref{Tq:finite1}--\ref{p_q1})
are subdominant and can be neglected. Let us consider first in
somewhat more detail the contribution $\propto \zeta^4$. This
contribution is the dominant one, for all $\nu<3/2$. It can be
easily seen to lead to a $w_q=1/3$ (since this contribution does
is not influenced by the trace of the stress energy tensor, its
contribution must be traceless), independent of $\epsilon$, and
therefore this contribution will always lead to a strong
backreaction, for larger enough values of $\epsilon$ (such that
$w_b$ is large). We do not believe however that this effect is due
to infrared particle production. The reason is the following. A
large backreaction, on physical grounds, is only expected when due
to particle creation the infrared becomes highly correlated. This
high amount of correlation leads to the infrared divergence and
thus is present only for $\nu>3/2$. Now not only does our
procedure not fix this contribution uniquely, in Appendix~C we
show that this contribution \emph{cannot} be determined uniquely
in the present model. The reason is that it turns out that the
contribution to the stress-energy tensor $\propto H^4\zeta^4$, is
ultraviolet divergent. The physical origin for this divergence is
probably the sudden matching between the two space-times we
consider. Such a divergence then needs to be renormalized, and
this renormalization inevitably introduces an arbitrariness in the
coefficient in front of the $H^4 \zeta^4$ term. Therefore we are
free to choose the constant $d_\nu$ and for the purpose of present
paper we choose the constant $d_\nu$ such that the term
proportional to $\zeta^4$ cancels. In this case we have that when
 $\nu<1/2$ a term proportional to $\zeta^2$ dominates, or
when $\nu>1/2$ the dominant term is proportional to
$\zeta^{1+2\nu-2n}$.

\subsubsection{The case when $\epsilon>1$, $0<\nu<1/2$}
\label{The case 3>epsilon>1, 0<nu<1/2}

 In this case the dominant contribution to the one loop
stress energy tensor comes from the term $\propto\zeta^2$
in~(\ref{Tq2:exact:dec}) and
\begin{eqnarray}
  T_q &\stackrel{\zeta\rightarrow \infty}{\longrightarrow}&
  \f{(1\!-\!6\xi)^2(1\!-\!\epsilon)(2\!-\!\epsilon)H^4}{16\pi^2}
        \zeta^2
\label{dec:Tq}
\\
  \rho_q &\stackrel{\zeta\rightarrow \infty}{\longrightarrow}&
  \f{(1\!-\!6\xi)^2(2\!-\!\epsilon)H^4}{64\pi^2}
        \big[-2\zeta^2\big]
\label{dec:rhoq}
\\
  p_q &\stackrel{\zeta\rightarrow \infty}{\longrightarrow}&
  \f{(1\!-\!6\xi)^2(2\!-\!\epsilon)H^4}{64\pi^2}
\Big[\f{2}{3}(1\!-\!2\epsilon)\Big]\zeta^2
\,,
\label{dec:pq}
\end{eqnarray}
from which we conclude:
\begin{eqnarray}
  w_q &\stackrel{\zeta\rightarrow \infty}{\longrightarrow}&
   - \f13 + \f23\epsilon = w_b + \f23 >w_b
\,,
\label{dec:wq}
\end{eqnarray}
where $w_b$ denotes the background equation of state
parameter~(\ref{w_b}). Notice that the condition, $\nu<1/2$
implies that $\xi<1/6$, if $\epsilon>2$ and $\xi>1/6$, if
$1<\epsilon<2$. In those regimes we therefore do not expect any
significant backreaction.

\subsubsection{The case when $\epsilon>1$, $\nu>1/2$}
\label{The case epsilon>1, nu>1/2}

 In this case the dominant contribution to $T_q$, $\rho_q$ and $p_q$
comes from a $\nu$ dependent power in
Eqs.~(\ref{Tq2:exact:dec}) and~(\ref{rhoq2:exact:dec})
({\it cf.} also Eqs.~(\ref{Tq:dec:nu-powers}--\ref{rhoq:dec:nu-powers})):
\begin{eqnarray}
  T_q &\stackrel{\zeta\rightarrow \infty}{\longrightarrow}&
  - \f{(1\!-\!6\xi)^2(1\!-\!\epsilon)(2\!-\!\epsilon)H^4}{16\pi^2}
   \bigg[\f{6\epsilon}{1\!+\!2\nu}-(5\epsilon\!-\!3)
       +(\epsilon\!-\!1)(2\nu\!+\!1)
   \bigg]\zeta^{1+2\nu}
\label{dec:Tq:2}
\\
  \rho_q &\stackrel{\zeta\rightarrow \infty}{\longrightarrow}&
  \f{(1\!-\!6\xi)^2(2\!-\!\epsilon)H^4}{64\pi^2}
   \bigg[\f{6\epsilon}{1\!+\!2\nu}-(5\epsilon\!-\!3)
       +(\epsilon\!-\!1)(2\nu\!+\!1)
   \bigg]\f{4\zeta^{1+2\nu}}{3-2\nu}
\label{dec:rhoq:2}
\\
  p_q &\stackrel{\zeta\rightarrow \infty}{\longrightarrow}&
  \f{(1\!-\!6\xi)^2(2\!-\!\epsilon)H^4}{64\pi^2}
   \bigg[\f{6\epsilon}{1\!+\!2\nu}-(5\epsilon\!-\!3)
       +(\epsilon\!-\!1)(2\nu\!+\!1)
   \bigg]\f43\Big[(\epsilon\!-\!1)+\f{1}{3\!-\!2\nu}\Big]\zeta^{1+2\nu}
\,,\quad
\nonumber\\
\label{dec:pq:2}
\end{eqnarray}
where we assumed that the term in square brackets does not vanish
and we took $\nu\neq 3/2$~\footnote{When $\nu=3/2$, $\rho_q$ and $p_q$
acquire a logarihmic contribution of $\zeta$, which yields the same
$w_q$ as is in equation~(\ref{dec:wq:2}).}.
From Eqs.~(\ref{dec:Tq:2}--\ref{dec:pq:2}) we conclude:
\begin{eqnarray}
  w_q &\stackrel{\zeta\rightarrow \infty}{\longrightarrow}&
   \epsilon - \f23 - \f23(\epsilon\!-\!1)\nu
    = w_b + \f13(\epsilon\!+\!1) - \f23(\epsilon\!-\!1)\nu
\,,
\label{dec:wq:2}
\end{eqnarray}
This equation implies that the quantum contribution will eventually dominate
provided,
\begin{equation}
 \nu > \f{\epsilon\!+\!1}{2(\epsilon\!-\!1)}
\,,
\label{dec:wq:2b}
\end{equation}
or equivalently when
\begin{eqnarray}
  \xi &<& - \f{\epsilon\!-\!1}{3(2\!-\!\epsilon)}\,,
 \qquad {\rm when}\;\;1<\epsilon<2
\nonumber\\
  \xi &>& \;\; \f{\epsilon\!-\!1}{3(\epsilon\!-\!2)}\,,
  \qquad\; {\rm when}\;\;\epsilon>2
\,.
\label{dec:wq:2c}
\end{eqnarray}
Together with the criterion $\xi<0$ when $0<\epsilon<1$, these
relations define the regions plotted in figure~\ref{fig5}, for
which one expects that the quantum one loop contribution to the
stress energy will eventually dominates over the classical stress
energy tensor that drives the Universe's expansion, and in which
case the quantum backreaction can influence -- and in fact change
-- the evolution of the Universe. These conclusions are correct,
provided the contributions~(\ref{dec:Tq:2}--\ref{dec:pq:2}) do not
vanish, which will be the case when the sum of the terms in square
brackets does not vanish, which is the case we consider next.

\subsubsection{The special case when $\epsilon>1$
and $\xi = -(\epsilon\!-\!1)/[3(2\!-\!\epsilon)]$ }
\label{The special case epsilon>1}

\begin{figure}
\begin{center}
\includegraphics[width=6in]{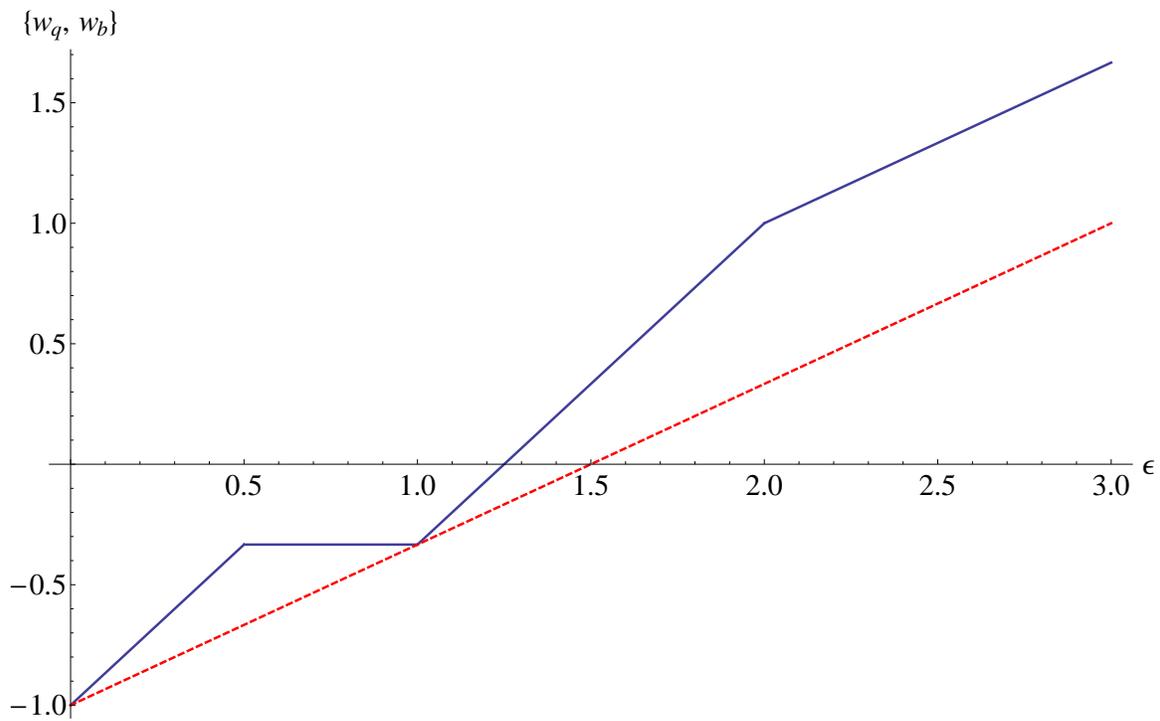}
\caption{$w_q$ (blue) and $w_b$ (red, dashed) versus $\epsilon$.
In this plot we choose $\xi$ to be on the boundary value which
seperates regimes where $w_q$ can become less then $w_b$ and
regimes where this can not happen. This implies $\xi=0$ for
$\epsilon<1$ and $\xi=-\f{\epsilon-1}{3(2-\epsilon)}$ for
$\epsilon>2$. $w_b$ is given by $w_b=-1+\f{2}{3}\epsilon$ and
$w_q$ is given by $w_q=-1+\f{4}{3}\epsilon$ $(0<\epsilon<1/2)$,
$w_q=-\f{1}{3}$ ($1/2<\epsilon<1$),
$w_q=-\f{5}{3}+\f{4}{3}\epsilon$ ($1<\epsilon<2$) and
$w_q=-\f{1}{3}+\f{2}{3}\epsilon$ ($\epsilon>2$). }\label{fig1}
\end{center}
\end{figure}

\begin{figure}
\begin{center}
\includegraphics[width=6in]{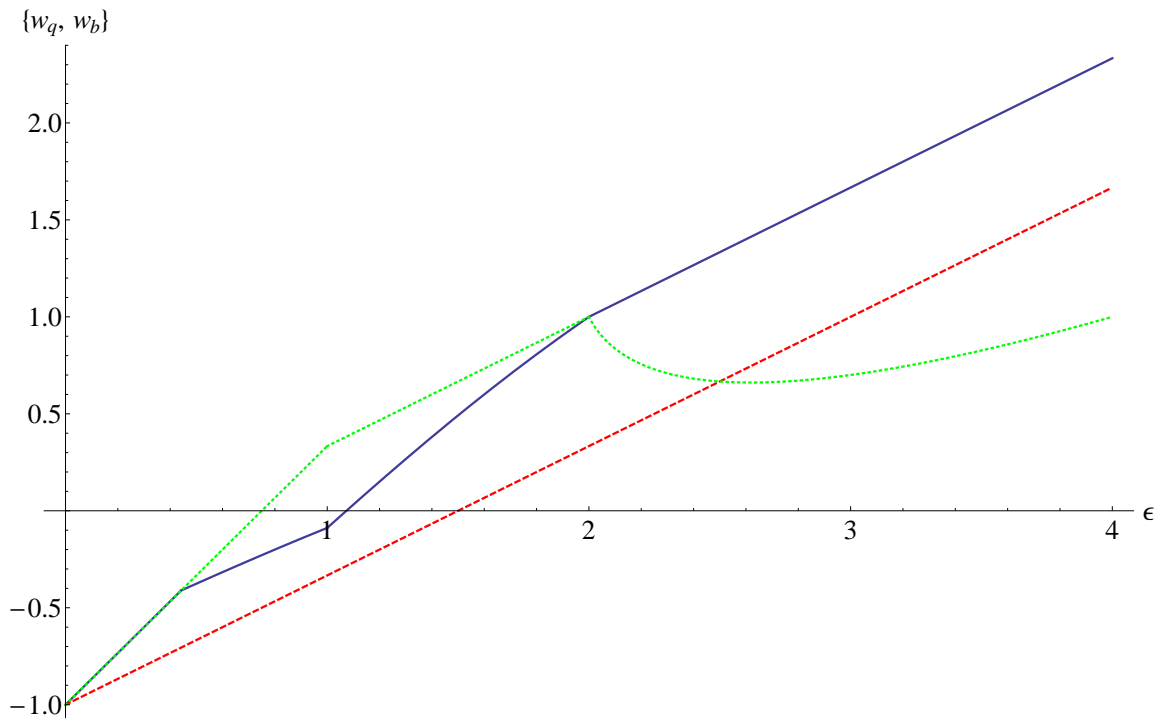}
\caption{$w_q$ and $w_b$ (red, dashed), versus $\epsilon$. This
plot shows positive values of $\xi$: $\xi=0.1$ (blue, solid) and
$\xi=1$ (green, dotted). The plotted values for $w_q$ are for the
$\xi=0.1$ curve: $w_q=-1+\f{4}{3}\epsilon$, for
$0<\epsilon<\epsilon_c$, where $\epsilon_c$ is determined by the
requirement $\nu=3/2$, implying $\epsilon_c=0.44\ldots$,
$w_q=-\f{2}{3}\nu+\f{\epsilon}{3}$ ($\epsilon_c<\epsilon<1$),
$w_q=\epsilon-\f{2}{3}(1+\nu)$, ($1<\epsilon<2$) and
$w_q=-\f{1}{3}+\f{2}{3}\epsilon$ ($\epsilon>2$). For the case
where $\xi=1$ we have $w_q=-1+\f{4}{3}\epsilon$ ($0<\epsilon<1$),
$w_q=-\f{1}{3}+\f{2}{3}\epsilon$ ($1<\epsilon<2$) and
$w_q=\epsilon-\f{2}{3}(1+\nu)$, ($\epsilon>2$) }\label{fig2}
\end{center}
\end{figure}

\begin{figure}
\begin{center}
\includegraphics[width=6in]{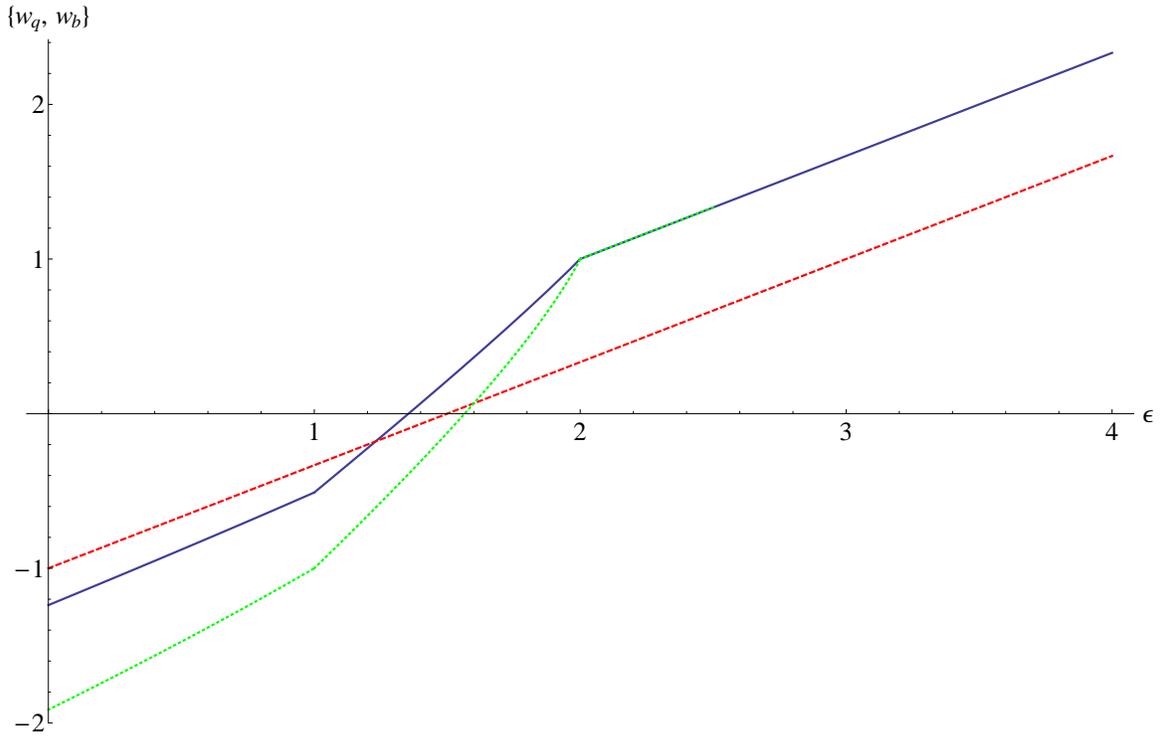}
\caption{$w_q$ and $w_b$ (red, dashed), versus $\epsilon$. This
plot shows negative values of $\xi$: $\xi=-0.1$ (blue, solid) and
$\xi=-0.5$ (green, dotted). The plotted values for $w_q$ are for
both curves $w_q=-\f{2}{3}\nu+\f{\epsilon}{3}$ ($0<\epsilon<1$),
$w_q=\epsilon-\f{2}{3}(1+\nu)$, ($1<\epsilon<2$) and
$w_q=-\f{1}{3}+\f{2}{3}\epsilon$ ($\epsilon>2$). }\label{fig3}
\end{center}
\end{figure}

\begin{figure}
\begin{center}
\includegraphics[width=6in]{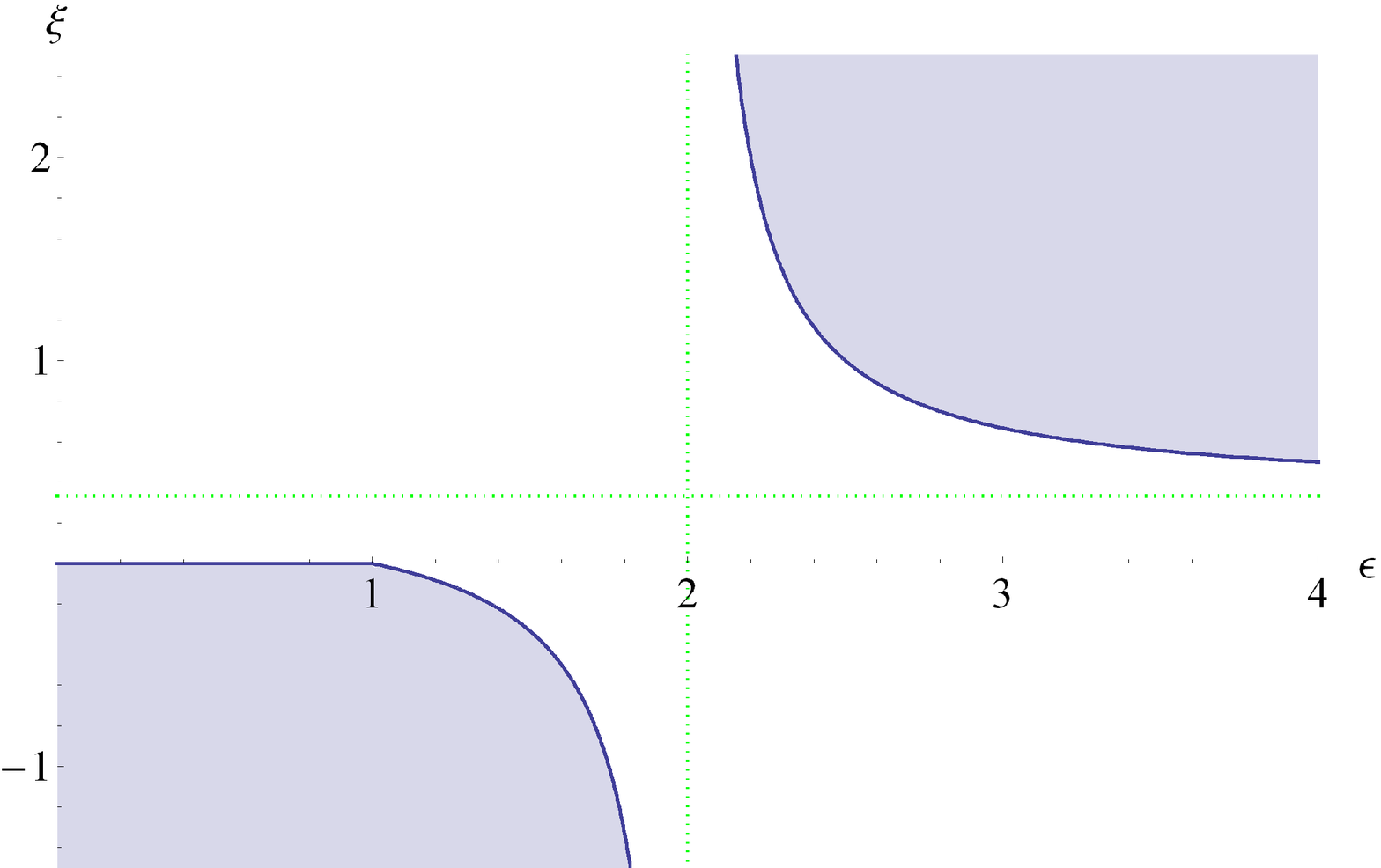}
\caption{Boundary values of $\xi$ as a function of $\epsilon$.
 The solid blue regions are the regions for which the quantum
stress energy scales slower that the classical stress energy, such
that the quantum contribution will eventually dominate over the
classical contribution. The regions are bounded by $\xi=0$
($0<\epsilon<1$) and  $\xi= -\f{\epsilon-1}{3(2-\epsilon)}$
($\epsilon>1$). The dotted asymptotes are given by $\epsilon=2$
and $\xi=1/3$.}\label{fig5}
\end{center}
\end{figure}

The case when  the sum of the terms in square
brackets~(\ref{dec:Tq:2}--\ref{dec:pq:2}) vanishes requires a special
attention, which is the case when,
\begin{equation}
  \nu\in\left\{1,\f{\epsilon\!+\!1}{2(\epsilon\!-\!1)}\right\}
\;\; \Longleftrightarrow  \;\;
\xi \in\left\{\f{(5\!-\!3\epsilon)(1\!+\!\epsilon)}{24(2\!-\!\epsilon)},
     -\f{\epsilon\!-\!1}{3(2\!-\!\epsilon)}\right\}
\,.
\label{nu,xi:cr}
\end{equation}
We shall only consider the second value, since we do not expect
anything significant for $\nu<3/2$. For this case the results of
the subsection~\ref{The case 3>epsilon>1, 0<nu<1/2} apply for all
$0<\nu<3/2$, and we have,

\begin{eqnarray}
  w_q &\stackrel{\zeta\rightarrow \infty}{\longrightarrow}&
  - \f13 +  \f23 \epsilon
  =  w_b + \f23 >w_b \qquad (0<\nu<3/2 \;\Leftrightarrow\; \epsilon>2)
\,,
\label{dec:wq:special}
\end{eqnarray}
When, on the other hand, $1<\epsilon<2$, such that $\nu>3/2$ and
$\xi<0$, the subleading terms in the
sums~(\ref{Tq:dec:nu-powers}--\ref{rhoq:dec:nu-powers}) dominate,
\begin{eqnarray}
  T_q &\stackrel{\zeta\rightarrow \infty}{\longrightarrow}&
  - \f{(1\!-\!6\xi)^2(1\!-\!\epsilon)^2(2\!-\!\epsilon)(5\!-\!3\epsilon)H^4}
      {16\pi^2}\zeta^{2/(\epsilon-1)}
\label{dec:Tq:3}
\\
  \rho_q &\stackrel{\zeta\rightarrow \infty}{\longrightarrow}&
  \f{(1\!-\!6\xi)^2(1\!-\!\epsilon)(2\!-\!\epsilon)(5\!-\!3\epsilon)H^4}
    {64\pi^2}
         \f{2(\epsilon\!-\!1)}{2\epsilon\!-\!3}\zeta^{2/(\epsilon-1)}
\label{dec:rhoq:3}
\\
  p_q &\stackrel{\zeta\rightarrow \infty}{\longrightarrow}&
  \f{(1\!-\!6\xi)^2(1\!-\!\epsilon)(2\!-\!\epsilon)(5\!-\!3\epsilon)H^4}
    {64\pi^2}
       \f{2(\epsilon\!-\!1)}{3(2\epsilon\!-\!3)}\Big[2(2\epsilon\!-\!3)+1\Big]
                  \zeta^{2/(\epsilon-1)}
\,.
\label{dec:pq:3}
\end{eqnarray}
This then implies for the equation of state parameter,
\begin{equation}
 w_q = -\f53+\f43\epsilon = w_b + \f23(\epsilon-1)
\qquad (\nu>3/2 \;\Leftrightarrow\; 1<\epsilon<2)
\,.
\label{dec:wq:special:2}
\end{equation}
Together with Eq.~(\ref{dec:wq:special:2}) this implies that
$w_q>w_b$, $\forall \epsilon>1$.\\
Let us now go back to the figures \ref{fig1}, \ref{fig2} and
\ref{fig3}, already discussed at the end of section \ref{Matching
onto acceleration: general}. Now we can also understand the part
of the plots for $\epsilon>1$. For the decelerating case we found
that quantum contributions can dominate if
$\xi<-\f{\epsilon-1}{3(2-\epsilon)}$ for $1<\epsilon<2$ and
$\xi>-\f{\epsilon-1}{3(2-\epsilon)}$ for $\epsilon>2$.
 In figure
\ref{fig1} we show the borderline case, thus for $\epsilon>1$ we
choose $\xi=-\f{\epsilon-1}{3(2-\epsilon)}$. This implies that for
$1<\epsilon<2$ we have that $w_q=-\f{5}{3}+\f{4}{3}\epsilon$,
conform (\ref{dec:wq:special:2}) and for $\epsilon>2$ we have
$w_q=-\f{1}{3}+\f{2}{3}\epsilon$, conform
(\ref{dec:wq:special}).
In figure \ref{fig2} we show $w_q$ for positive values of $\xi$. A
positive value of $\xi$, less then 1/6 means that $\nu>1/2$ for
$\epsilon<2$ and $\nu>1/2$ for $\epsilon>2$ thus in this case we
show $w_q=\epsilon-\f{2}{3}-\f{2}{3}(\epsilon-1)\nu$, conform
(\ref{dec:wq}), for the region $1<\epsilon<2$ and
$w_q=-\f{1}{3}+\f{2}{3}\epsilon$, confrom (\ref{dec:wq:2}) for
$\epsilon>2$. On the other hand if $\xi>1/6$, we have that for
$1<\epsilon<2$ that $\nu<1/2$ and for $\epsilon>2$, $\nu>1/2$.
Thus in this case it is precisely the other way around. Moreover ,
since now $\xi>\f{\epsilon-1}{3(\epsilon-2)}$ in the region where
$\epsilon>2$, we find conform (\ref{dec:wq:2c}) that in this
regime $w_q$ can become smaller then $w_b$.
Finally in figure \ref{fig3} we show $w_q$ for negative values of
$\xi$. If $\xi$ is negative, we always have that $\nu>1/2$ for
$1<\epsilon<2$ and $\nu<1/2$ for $\epsilon>2$. And thus we show
$w_q=\epsilon-\f{2}{3}-\f{2}{3}(\epsilon-1)\nu$, conform
(\ref{dec:wq}), for the region $1<\epsilon<2$ and
$w_q=-\f{1}{3}+\f{2}{3}\epsilon$, confrom (\ref{dec:wq:2}) for
$\epsilon>2$. We find, conform (\ref{dec:wq:2c}) that in the
region $1<\epsilon<2$ we can have that $w_q<w_b$.\\
This completes our discussion for now. However there are special
points, corresponding to $\nu=3/2$ and $\nu=5/2$, where the above
discussion fails because of the logarithms appearing in $\rho_q$ and $p_q$.
However it turns that the scaling of the quantum
corrections, $w_q$ does not deviate from the behavior described
above. Details on this we present in Appendix~C.

\section{Comparison with the cut-off regulated $T_{\mu\nu}$}
\label{s_cutoff}

In earlier work~\cite{Janssen:2008px} we have calculated the one
loop stress energy due to scalar field fluctuations, where the
infrared was regulated, by working on a compact spatial manifold,
or in other words, by assuming that the Universe is a comoving box, with
periodic boundary conditions (which to a good approximation can be
described by a comoving infrared momentum cut-off).

 Generalising the result from~\cite{Janssen:2008px} to a nonminimally coupled
scalar, we obtain for the renormalised stress-energy tensor
\begin{equation}\label{Tmnfinal}
    \begin{split}
        \langle
        0|T_{\mu\nu}|0\rangle=&-\f{\epsilon(2-\epsilon)(1-6\xi)^2}{16\pi^2}H^4
  \bigg[\gamma_E+\ln\Big(\f{(1-\epsilon)^2
        H^2}{4\pi\mu^2}\Big)+\psi(\f{1}{2}-\nu)+\psi(\f{1}{2}+\nu)\\
        &+2\f{1-2\nu'}{1-2\nu}+2\f{1+2\nu'}{1+2\nu}\bigg]
   \Big(\epsilon
        a^2 \delta_\mu^0\delta_\nu^0+(\epsilon-\f{3}{4})g_{\mu\nu}\Big)\\
        &+\f{H^4(1-6\xi)}{16\pi^2}
  \bigg[\Big(-(1-4\xi)(2-\epsilon)\epsilon^2-\f{2}{3}(1-6\xi)(1-4\epsilon+\epsilon^2)\epsilon\Big)a^2\delta_\mu^0\delta_\nu^0\\
        &+\Big(-\f{1}{8}\big(-7+8\epsilon(1-4\xi)+30\xi\big)-\f{1-6\xi}{6}\big(3-22\epsilon+22\epsilon^2-4\epsilon^3\big)\Big)g_{\mu\nu}\bigg]\\
        &+\sum_{N=0}^\infty \left(\langle\Omega|T_{\mu\nu}|\Omega\rangle_{N}
            +\langle\Omega|T_{\mu\nu}|\Omega\rangle^{N}\right)
\qquad
   \end{split}
\end{equation}
where $\nu^\prime = d\nu/dD|_{D=4}=$ and the terms in the last
line are the infrared corrections arising from the infrared
cutoff, which can be written as~\cite{Janssen:2008px}
\begin{equation}
\label{Tmn:N<}
    \begin{split}
\langle0|T_{\mu\nu}|0\rangle_{N} &= (\rho_N+p_N)
a^2\delta_\mu^0\delta_\nu^0 + p_N g_{\mu\nu}\\
&=\f{1}{4\pi^{5/2}}\f{\big(2(1-\epsilon)(N-\nu)+3-\epsilon\big)}{3+2N-2\nu}\f{\Gamma(\nu-N)\Gamma(2\nu-N)}{\Gamma(\f{1}{2}+\nu-N)\Gamma(N+1)}\\
&\qquad\times H^4(1-\epsilon)^2z_0^{2N+3-2\nu}(1-6\xi)(N-\nu)\\
&\qquad\times\bigg[\f{1}{3}\Big(\f{4}{1-2N+2\nu}-(1-\epsilon)\Big)
 a^2\delta_\mu^0\delta_\nu^0+\f{1}{3}\Big(\f{1}{1-2N+2\nu}-(1-\epsilon)\Big)
 g_{\mu\nu}\bigg]
\end{split}
\end{equation}
and
\begin{equation}
\label{Tmn:N>}
    \begin{split}
\langle0|T_{\mu\nu}|0\rangle^{N}
&=\f{1}{4\pi^{5/2}}\f{\big(2(1-\epsilon)(N+\nu)+3-\epsilon\big)}{3+2N+2\nu}\f{\Gamma(-\nu-N)\Gamma(-2\nu-N)}{\Gamma(\f{1}{2}-\nu-N)\Gamma(N+1)}\\
&\qquad\times H^4(1-\epsilon)^2z_0^{2N+3+2\nu}(1-6\xi)(N+\nu)\\
&\qquad\times\bigg[\f{1}{3}\Big(\f{4}{1-2N-2\nu}-(1-\epsilon)\Big)
    a^2\delta_\mu^0\delta_\nu^0
  +\f{1}{3}\Big(\f{1}{1-2N-2\nu}-(1-\epsilon)\Big)g_{\mu\nu}\bigg]
\,,
\end{split}
\end{equation}
where $z_0 = k_0|\eta|$, $k_0$ is the (comoving) infrared cut-off scale
($k_0=2\pi/L$, where $L$ is the comoving size of the Universe) and $\eta$
is conformal time, defined by $a\eta=-1/[(1-\epsilon)H]$.
We assume that the size of the Universe is at an initial time $\eta_0$
super-Hubble, such that $k_0|\eta_0| = k_0/[(1-\epsilon)a_0H_0]\ll 1$.
Notice that in an accelerating
universe ($0<\epsilon<1$) at late times
$z_0$ decreases, such that $z_0\rightarrow 0$ as $\eta\rightarrow 0-$
(or equivalently $a\rightarrow \infty$). On the other hand, in a decelerating
universe $\eta$ increases, resulting in an increasing $z_0$.
When $z_0\sim 1$, the size of the Universe
becomes comparable to the Hubble radius. Even later
the comoving box shrinks to sub-Hubble sizes, and $z_0\rightarrow \infty$
when $\eta, a\rightarrow \infty$. This case requires a special attention,
and it is treated below.

 We shall now construct the leading order contribution from the
corrections~(\ref{Tmn:N<}--\ref{Tmn:N>}) to the stress energy
tensor in both accelerating and decelerating universes. If $z_0$
approaches zero, we see that the leading order contribution comes
from the $N=0$ term of (\ref{Tmn:N<}). This term is growing if
$\nu>3/2$, as is expected, since this is the requirement for an
infrared divergence~(\ref{IRdiv:BD}). The case $\nu=3/2$ leads to
a logarithmic growth which, for brevity, we do not study here. The
results turn out to be analogous to the case we consider here,
described in Eqs.~(\ref{spec:nu32:acc:Tq}--\ref{spec:nu32:acc:pq})
for the accelerating case. If $\nu>3/2$, the leading contribution
in an accelerating universe is
\begin{equation}\label{cutoff_LO}
    \begin{split}
    \langle 0 |T_{\mu\nu}|0\rangle
  \;\stackrel{z_0\rightarrow 0}{\longrightarrow}\;
 &\f{4
    H^4(1-\epsilon)^4}{3\pi^2(1+2\nu)(2\nu-3)}\Big(\f{z_0}{2}\Big)^{3-2\nu}(1-6\xi)\Gamma(\nu)\Gamma(\nu+1)\Big(\f{3-\epsilon}{2(1-\epsilon)}-\nu\Big)\\
    &\times\,\bigg[\Big(\f{3+\epsilon}{2(1-\epsilon)}-\nu\Big)a^2\delta_\mu^0\delta_\nu^0+\Big(\f{\epsilon}{2(1-\epsilon)}-\nu\Big)g_{\mu\nu}\bigg]
  +\mathcal{O}(z_0^{5-2\nu})
\,.
\end{split}
\end{equation}
This expression vanishes when $\xi=0$, since then
$\nu=(3-\epsilon)/[2(1-\epsilon)]$. In that specific case we need
the next to leading order contribution, which is proportional to
$z_0^{5-2\nu}$. The form of the leading order
contribution~(\ref{cutoff_LO}) very similar to corresponding
matching case~(\ref{acc:Tq:2}--\ref{acc:wq:2}), with
$z_0\rightarrow \zeta$ (the precise numerical coefficients
multiplying the leading order terms are, of course, different).
Thus in an accelerating universe, the two regularization
procedures are qualitatively the same. The reasons for this is
that any infrared regularization implies that we effectively
suppress modes with wavelengths larger then some scale, be it
given by $z_0$ or $\zeta$. In an accelerating universe, this scale
grows faster then the Hubble radius and therefore the precise
details of this regularization become less and less visible as
time goes on. Therefore we indeed find that at late enough times
the two regularization schemes give qualitatively the same
result.

However, as mentioned above, in a decelerating space-time we have
a different situation: the sums over $N$ run to infinity and,
since $z_0$ grows, this leads in principle to fast growing terms.
Moreover we see that this happens when the Universe's size becomes
sub-Hubble, independent of $\nu$. Thus also infrared perfectly
finite space-times become dominated by the cut-off. The most reasonable
approach is is this case to sum the sums over $N$ when $z_0\ll 1$ and then
analytically extend to late times when $z_0\gg 1$ and the
Universe's size is sub-Hubble. We shall now illustrate how this
procedure works by calculating the trace of the stress energy
tensor. After taking the trace of (\ref{Tmn:N<}) we can perform
the sum over $N$ to obtain
\begin{equation}
    \begin{split}
    \sum_{N=0}^{\infty} \langle 0
    |T^\mu{}_{\mu}|0\rangle_N&=\f{2\Gamma(\nu)^2}{\pi^3(5-2\nu)(3-2\nu)}(1-6\xi)H^4(1-\epsilon)^4\Big(\f{z_0}{2}\Big)^{3-2\nu}\\
    &\qquad\times\Bigg[\nu(5-2\nu)\Big(\f{3-\epsilon}{2(1-\epsilon)}-\nu\Big){}_2
    F_3\Big(\f{1}{2}-\nu,\f{3}{2}-\nu;1-2\nu,\f{5}{2}-\nu,-\nu;-z_0^2\Big)\\
    &\qquad+\f{z_0^2}{2}(3-2\nu){}_2
    F_3\Big(\f{3}{2}-\nu,\f{5}{2}-\nu;2-2\nu,\f{7}{2}-\nu,1-\nu;-z_0^2\Big)\Bigg]\,.
    \end{split}
\end{equation}
The leading order can be studied by considering the asymptotic
expansion of the ${}_2 F_3$ hypergeometric functions. We have in general
\begin{equation}
    \begin{split}
        {}_2
        F_3&\Big(a_1,a_2;b_1,b_2,b_3;-z\Big)=\f{\Gamma(b_1)\Gamma(b_2)\Gamma(b_3)}{\Gamma(a_1)\Gamma(a_2)}\Bigg\{\f{\Gamma(a_1)\Gamma(a_2\!-\!a_1)}{\Gamma(b_1\!-\!a_1)\Gamma(b_2\!-\!a_1)\Gamma(b_3\!-\!a_1)}z^{-a_1}\Big(1+\mathcal{O}\big(z^{-1}\big)\Big)\\
        &+\f{\Gamma(a_2)\Gamma(a_1-a_2)}{\Gamma(b_1-a_2)\Gamma(b_2-a_2)\Gamma(b_3-a_2)}z^{-a_2}\Big(1+\mathcal{O}\big(z^{-1}\big)\Big)\\
        &+\f{z^\chi}{\sqrt{\pi}}\bigg[\cos\Big(\pi\chi+2\sqrt{z}\Big)
    +\f{1}{16\sqrt{z}}\Big((3a_1+3a_2+b_1+b_2+b_3-2)(8\chi-2)\\
        &\qquad\qquad+16(b_1b_2+b_1b_3+b_2b_3-a_1a_2)-3\Big)\sin\Big(\pi\chi+2\sqrt{z}\Big)\bigg]\Big(1+\mathcal{O}\big(z^{-1}\big)\Big)\Bigg\},
    \end{split}
\end{equation}
with
\begin{equation}
\chi=\f{1}{2}\Big(a_1+a_2-b_1-b_2-b_3+\f{1}{2}\Big).
\end{equation}

Using this we find that the leading order terms are
\begin{equation}\label{decel_trace}
    \begin{split}
    \sum_{N=0}^{\infty} \langle 0
    |T^\mu{}_{\mu}|0\rangle_N&=
    \f{(1\!-\!6\xi)(1\!-\!\epsilon)^3H^4}{8\pi^2\sin^2(\pi\nu)}
         \bigg[\f{z_0^2}{2}
  + z_0^2\Big(\f{5\epsilon\!-\!1}{4}+(1\!-\!\epsilon)\nu^2\Big)
              \sin(2z_0\!+\!\pi\nu)
\\
    &\hskip 3.8cm
  -z_0^3(1\!-\!\epsilon)\cos(2z_0\!+\!\pi\nu)\bigg]
  +\mathcal{O}\big(z_0^0\big)
\,.
    \end{split}
\end{equation}
 The second series $\sum_{N=0}^{\infty} \langle 0 |T^\mu{}_{\mu}|0\rangle^N$
in Eq.~(\ref{Tmnfinal}) can be obtained
from~(\ref{decel_trace}) by interchanging $\nu$ with $-\nu$
and can be easily added, resulting in,
\begin{equation}
    \begin{split}
  T_q \simeq \sum_{N=0}^{\infty} \Big(\langle 0|T^\mu{}_{\mu}|0\rangle_N
     + \langle 0|T^\mu{}_{\mu}|0\rangle^N\Big)
   &=\f{(1\!-\!6\xi)(1\!-\!\epsilon)^3H^4}{8\pi^2\sin^2(\pi\nu)}
\\
    &\hskip -6.3cm
 \times\,\bigg[z_0^2
  + z_0^2\bigg(\f{5\epsilon\!-\!1}{2}+2(1\!-\!\epsilon)\nu^2\bigg)
              \cos(\pi\nu)\sin(2z_0)
  -2z_0^3(1\!-\!\epsilon)\cos(\pi\nu)\cos(2z_0)\bigg]
  +\mathcal{O}\big(z_0^0\big)
\,.
    \end{split}
\quad
\label{decel_trace:2}
\end{equation}
Selecting the leading order term in~(\ref{decel_trace:2})
we can recast it as,
\begin{equation}
 T_q \simeq \f{(1\!-\!6\xi)(\epsilon\!-\!1)H_0k_0^3}{4\pi^2}
       \f{\cos(\pi\nu)}{\sin^2(\pi\nu)}
     \bigg[\cos\bigg(\f{2k_0}{(\epsilon\!-\!1)Ha}\bigg)
         + {\cal O}(a^{1\!-\!\epsilon})\bigg]a^{-\epsilon-3}
\,.
\label{decel_trace:3}
\end{equation}
We shall not attempt to evaluate $\rho_q$ and $p_q$ in this case, since
the asymptotic expansion would result in an expression that depends on
the lowest value of $z_0$. Instead we shall compare the corresponding traces.
When Eq.~(\ref{decel_trace:3}) is compared with the background contribution,
$T_b = \rho_b-3p_b\propto 1/a^{3(1+w_b)}=a^{-2\epsilon}$,
one finds that
\begin{equation}
   \f{T_q}{T_b} \sim \f{k_0^3}{H_0M_P^2}a^{\epsilon-3}
              \cos\bigg(\f{2k_0}{(\epsilon\!-\!1)Ha}\bigg)
\,,
\label{decel_trace:4}
\end{equation}
which grows when $\epsilon>3$. Thus we find that the limiting case
is kination, for which $\epsilon=3$ and $\rho_b\propto 1/a^6$, and
the quantum contribution to the trace $T_q$ given
in~(\ref{decel_trace:3}) scales the same as the background
contribution to the trace. For all $\epsilon>3$ we thus find that
the quantum contribution will eventually dominate over the
background energy density. There is one exception: as long as
$\nu$ is not half integer, the scaling~(\ref{decel_trace:4}) is
correct. However when $\nu$ is half integer, $\cos(\pi\nu)=0$ and
the leading contribution~(\ref{decel_trace:3}) vanishes. The
dominant contribution is then the term $\propto z_0^2$ in
Eq.~(\ref{decel_trace:2}). We have seen in section \ref{The case
3>epsilon>1, 0<nu<1/2} that such a contribution will never lead to
a strong backreaction. This is in contrast with what we found in the
present work. Provided that we correctly disregarded the $\zeta^4$
term, which arose as an integration constant, we have found in
section \ref{Matching onto deceleration: general} that, if the
infrared is regulated using the mode matching, for all $\nu>1/2$,
the scaling of the quantum energy is indeed governed by $\nu$ and
not $\epsilon$.

\section{Summary and discussion}\label{Summary and discussion}
In order to facilitate the reading of this rather technical paper,
we shall now recap our main results. The main motivation for this
paper is to study the role of the quantum infrared fluctuations
generated by the expansion of the Universe for massless scalar
fields, with a possible coupling to the Ricci scalar.

It is well known that the infrared (IR) sector of such a scalar
field posses problems on a cosmological background: in space times
with a negative pressure, and constant acceleration/deceleration
parameter $\epsilon=-\dot H/H^2$, the Bunch-Davies (BD) vacuum of
massless minimally coupled scalars is infrared
singular~\cite{Vilenkin:1982wt} (see section \ref{s_scalar}).

Of course, we know that our Universe must be infrared finite. That
means that the infrared sector of the theory must be regulated.
It is currently unknown what is the precise nature of the
regularisation scheme realized in our Universe. Hence it is worth
investigating different regularisation schemes and compare the
results. In the end however, the presence of this infrared
singularity is precisely what makes the present study interesting.
The singularity indicates that there is a growth in long range
correlations due to particle production. Even after regulating the
infrared divergence, this growth of the correlations is still
there, since it is simply a physical effect. These growing
correlations can then in principle  -- after a sufficient amount
of time -- contribute significantly to the energy density in the
Universe. If this is so, they might change the evolution of the
Universe significantly.


 In this work we focus our attention on the regularisation of the infrared which is
executed by an epoch of the very early universe which has very
little particle production and whose BD vacuum is therefore
infrared finite. For definiteness we choose this early epoch to be
radiation era. We then match it onto a constant $\epsilon$ space
time, and calculate the corresponding coincident scalar
propagator.
The exact solution for the mode functions can be expressed in
terms of Hankel functions~(\ref{mode_sol}) with a rather
complicated index $\nu$~(\ref{nu}). The BD vacuum corresponds to
the choice~(\ref{BD vacuum}), implying that one considers only
positive frequency modes. As can be seen from Eq.~(\ref{IRdiv:BD})
the propagator for the BD vacuum is IR singular whenever $\nu\geq
(D-1)/2$, where $D$ denotes the number of spacetime dimensions. At
the matching we require continuity of both the mode functions and
their first derivatives. This implies that after the matching, the
mode functions become a mixture of positive and negative frequency
modes, in such way that the (coincident) propagator~(\ref{int}) is
IR finite for all $\epsilon$. The propagator still suffers from
the standard (logarithmic) ultraviolet divergence, which in
dimensional regularisation appears as a $1/(D-4)$
divergence~(\ref{propUV}). This divergence induces an analogous
divergence in the stress energy tensor~(\ref{Tq:div}), and can be
removed using standard techniques by an $R^2$
counterterm~(\ref{ct:trace}).

To construct the propagator, we have first calculated the
coincident propagator for half integer
$\nu$~(\ref{propIR}--\ref{propIR:2}) and then analytically
extended the result to all (complex)
$\nu$~(\ref{prop:full:resum}).
 Since the coincident propagator
for half integers~(\ref{prop:full}) is valid for all times later
than the matching, we suspect that the analytic
extension~(\ref{prop:full:resum}) is unique.
From the coincident propagator, one can derive the trace of the
one loop stress energy tensor $T_q$, based on
equation~(\ref{Tmn:trace}). The symmetries of the background space
time dictate the perfect fluid form of the quantum contribution to
the stress energy tensor and using the covariant conservation of
the stress energy tensor ~(\ref{conservation}), one can derive the
individual components (quantum energy density $\rho_q$ and
pressure $p_q$) of the one loop $(T_{\mu\nu})_q$.
The general results of this procedure are for accelerating
universe presented in Eqs.~(\ref{Tq2:exact}) and in
Eqs.~(\ref{rhoq2:exact}--\ref{pq2:exact}) and for decelerating
universes in Eqs.~(\ref{Tq2:exact:dec}--\ref{rhoq2:exact:dec}).
Since we
obtain $\rho_q$ and $p_q$ by an integration procedure, our results
are unique up to an integration constant. Adding this integration
constant corresponds to adding a component to $\rho_q$ and $p_q$,
which scales as a radiation fluid, $\propto 1/a^4$, and thus does
not contribute to the trace of the stress energy tensor $T_q$. To
fix this component uniquely, we would need to know the propagator
away from the coincidence limit, which we have not derived in this
paper. Since this undetermined radiation component yields a
subdominant contribution at late times for accelerating universes
and thus it is of no relevance for our discussion of the late time
quantum stress energy tensor in section~\ref{Matching onto
acceleration: general}. It may be however relevant in decelerating
universes. For the purpose of this work we have ignored this
radiation contribution in our analysis of the late time quantum
stress energy tensor in decelerating universes presented in
section~\ref{Matching onto deceleration: general}.

In order to study the significance of the quantum one loop stress
energy tensor, in subsections~\ref{Matching onto acceleration:
general} and~\ref{Matching onto deceleration: general} we study in
detail
the equation of state parameter $w_q = p_q/\rho_q$ both in
accelerating and in decelerating universes. If the following
criterion is satisfied~(\ref{w_q-w_b}),
\begin{equation}
 w_q<w_b
\,, \nonumber
\end{equation}
(where $w_b=p_b/\rho_b$ is the equation of state parameter of the
background fluid, driving the expansion of the universe), then the
quantum contribution to the energy density dominates over the
background contribution at late times. In this work we do not
attempt to analyze what exactly happens in such a case (to start
with, the assumptions underlying our calculations would become
incorrect), but simply want to answer the question \emph{if} and
under what conditions this criterion is ever satisfied.\\
Our results are presented in detail in sections~~\ref{Matching
onto acceleration: general} and~\ref{Matching onto deceleration:
general}. Since several cases require a separate discussion, it is
difficult to get a quick grasp of the results. In order to
facilitate a quicker understanding of our results we present in
figures~\ref{fig1}--\ref{fig3} the quantum equation of state
parameter $w_q$ {\it vs} the background equation of state
parameter $w_b=-1+(2/3)\epsilon$. The criterion~(\ref{w_q-w_b})
then tells us when the quantum contribution becomes dominant over
the background contribution at late times. We shall now describe
our results in some detail.

 Firstly, in figure~\ref{fig5} the shaded regions represent the regions
in parameter space $\{\xi,\epsilon\}$ where the criterion
$w_q<w_b$ is satisfied. For accelerated universes ($0<\epsilon<1$)
this means simply that when $\xi<0$, $w_q<w_b$. For decelerating
universes ($\epsilon>1$) we need to distinguish two cases. When
$1<\epsilon<2$, then
$\xi<-(\epsilon-1)/[3(2-\epsilon)]$~(\ref{dec:wq:2c}) assures
$w_q<w_b$. When on the other hand $\epsilon>2$,
$\xi>(\epsilon-1)/[3(\epsilon-2)]$~(\ref{dec:wq:2c}) is required
in order that $w_q<w_b$. Keeping in mind that the Ricci scalar
curvature $R=6(2-\epsilon)H^2$ is positive when $0<\epsilon<2$ and
negative when $\epsilon>2$, we see that $w_q<w_b$ can be met only
when the effective scalar `mass' parameter $m_{\rm eff}^2=\xi R$
is (sufficiently) negative. A careful analysis presented in
subsections~\ref{The special case when epsilon<1, xi=0}
and~\ref{The case epsilon>1, nu>1/2} shows that at the boundary of
the shaded region in all cases $w_q\geq w_b$ is satisfied, such
that the quantum contribution can never become important. This can also be seen from figure \ref{fig1}.\\
In figure~\ref{fig2} we show $w_q$ {\it vs} $w_b$ for positive
$\xi>0$. When the coupling $\xi$ is smaller than 1/3, we have that
$w_q>w_b$ such that the quantum stress energy can never be
important. When however $\xi> 1/3$, in spacetimes with
$\epsilon>2$, $w_q$ can become smaller than $w_b$, indicating the
late time dominance of the one loop contribution. \\
Finally, if $\xi$ is negative, which is shown in
figure~\ref{fig3}. In this case $w_q<w_b$ is always met for
accelerating cases, but also for decelerating cases whenever
$\xi<\xi_{\rm cr}=-(\epsilon-1)/[3(2-\epsilon)]$. Notice that the
critical $\epsilon$ (for which $\xi=\xi_{\rm cr}$) is always
smaller than $2$.

 Apart from the regularisation scheme presented in this work that involves matching from
a nonsingular spacetime, other infrared regularisation schemes are
possible. We do not know which regularisation scheme was chosen by
our Universe. We do know however that, since infrared
regularisation is physical,
 there are in principle physical observables that can distinguish between
different regularisation schemes, and thus we should be able (at
least in principle) decide which one was picked by our Universe.
But in order to find out the answer to that question, we need to
investigate different plausible IR regularlisation schemes. Other
regularisation schemes include: placing the Universe in a large
(comoving) box (this corresponds to an infrared momentum cutoff)
which has been explored in
Refs.~\cite{TW3,Janssen:2008px,Janssen:2009pb}, a tiny scalar mass
(it is not clear whether that is possible to implement for the
graviton), a positive spatial curvature, subtracting an
adiabatic~\cite{Parker:2007ni} or a comoving
vacuum~\cite{Agullo:2008ka,Agullo:2009vq}, {\it etc}. While most
of these schemes are physically well motivated, the implementation
is often hindered by our lack of knowledge of the relevant
propagators. In particular, we know the propagators in positively
curved universes and for massive fields only in very special cases
(de Sitter space, radiation era), which is not enough to conduct a
sufficiently general analysis of the quantum backreaction
regulated this way.

 Owing to the fact that the analysis of the Universe in a finite comoving box
can be well approximated by an infrared momentum cutoff, in
Ref.~\cite{Janssen:2008px} we were able to perform the one loop
analysis of the massless scalar backreaction in expanding
universes with a constant $\epsilon$ parameter. In
section~\ref{s_cutoff} we present in some detail a comparison
between the two regularisation schemes. We find that in
accelerated universes the leading order contributions to the
corresponding late time stress energy tensors are of a similar
form in both regularisation schemes in the sense that the leading
order behavior with the scale factor is identical ({\it cf.}
Eqs.~(\ref{acc:Tq:2}--\ref{acc:wq:2}) and~(\ref{cutoff_LO})),
albeit the coefficients of the leading order terms differ. In
fact, the coefficients are comparable when the cut-off scale $k_0$
is chosen to be equal to the horizon scale $\hat H$ at the
matching, {\it i.e.} when the initial comoving size of the
Universe is of the order the Hubble radius. This is
understandable, given the fact that the comoving box in
accelerating universes grows with respect to the Hubble radius,
{\it i.e.} as times goes on the Universe's size becomes more and
more super-Hubble. The precise details about what precisely
happens on super-Hubble scales become less and less `visible' as
time goes on. This is precisely what we find: at late times the
two approaches give qualitatively the same answer.

 On the other hand, in decelerating space-times the matching
and cut-off regularisation schemes yield very different results.
 The leading order contributions to the
trace of the stress energy tensor are in this case ({\it cf.}
Eqs.~(\ref{dec:Tq:2}) for $\nu>1/2$, Eq.~(\ref{dec:Tq}) for
$\Re[\nu]<1/2$ and~(\ref{decel_trace:2}--\ref{decel_trace:3}))
\begin{eqnarray}
        \langle 0| T^{\mu}{}_\mu|0\rangle &\propto& H^4 z_0^3
     \cos(2z_0)+\mathcal{O}(z_0^2)\qquad\! ;\qquad ({\rm cut-off})
\\
   \langle 0| T^{\mu}{}_\mu|0\rangle &\propto&
       \begin{cases}
             H^4 \zeta^{2\nu+1}
              +\mathcal{O}(\zeta^{2\nu-1})\quad &;\qquad
                         ({\rm matching},\; \nu>1/2) \cr
      H^4 \zeta^{2} +\mathcal{O}(\zeta^{2\nu+1},\zeta)\quad &;
                       \qquad ({\rm matching},\; \Re[\nu]<1/2), \cr
      \end{cases}
\label{comparison:box-match:decel}
\end{eqnarray}
where $\zeta = \hat a\hat H/(aH)$ and $z_0=k_0/[(\epsilon-1)aH]
     = \{k_0/[(\epsilon-1)\hat a\hat H]\}\zeta$, such that
$z_0/\zeta$ is a constant given by the ratio of the physical
cutoff scale $k_0/\hat a$ and the (inverse) particle horizon $\sim
(\epsilon-1)\hat H$ at the matching time. For the cut-off
regulated case we find that the growth is independent of $\nu$.
The quantum contribution in this case becomes more and more
dominant if $\epsilon$ becomes larger. In the mode-matching case
we see that the growth is dependent on $\nu$ and moreover the
effect is more profound the greater $\nu$ is. This is exactly what
one would expect based on~(\ref{prop:coincident div}). What
happens physically is that, in a decelerating space-time, physical
scales grow slower then the Hubble radius, and eventually the
comoving size of the Universe becomes sub-Hubble. This would
eventually dominate all effects and thus what one is looking at
then has nothing to do anymore with infrared particle production!
The results~(\ref{comparison:box-match:decel})
and~(\ref{decel_trace:2}--\ref{decel_trace:3}) are obtained by
analytically extending to the limit $z_0\rightarrow \infty$, which
is precisely the limit in which the Universe's size becomes
sub-Hubble. Thence, not surprisingly, in this limit $T_q\propto
1/V_c$ becomes a function of the comoving volume $V_c$ of
space-time, representing an observable indicating the size of the
Universe. In the infrared reularization presented in this work,
all potentially relevant effects are however due to infrared
particle production, and since this is the physical effect we wish
to study, we feel that this approach therefore is advantageous in
decelerating spacetimes.

\section*{Acknowledgements}

We would like to thank Richard P. Woodard for his contributions at
the early stages of this project and useful suggestions later on
and Shun Pei Miao for critically reading the manuscript.

\section*{Appendix A: Acting with the d'Alembertian}
\label{Appendix A}

In order to calculate the contribution to
the trace of the stress energy tensor~(\ref{Tmn:trace}),
the following identities are useful,
\begin{eqnarray}
 \square [H^2 f(\zeta)] &=& (1\!-\!\epsilon)H^4
     \Big[
          6\epsilon
        + (3\!-\!5\epsilon)\zeta\f{d}{d\zeta}
        - (1\!-\!\epsilon)\zeta\f{d}{d\zeta}\zeta\f{d}{d\zeta}
     \Big]f(\zeta)
\label{box identity 1}
\\
 \square [H^2 \ln(H^2)] &=& 2\epsilon H^4
     \Big[3(1\!-\!\epsilon)\ln(H^2)
        + (3\!-\!5\epsilon)
     \Big]
\,,
\label{box identity 2}
\end{eqnarray}
where we made use of $\square = -(\partial_t + 3H)\partial_t$ and
$\partial_t = -(1\!-\!\epsilon)H\zeta\partial_\zeta$.
 From these we easily get the following useful identities,
\begin{eqnarray}
 \square [H^2 \ln(1\pm \zeta)] &=& (1\!-\!\epsilon)H^4
     \Big[
          6\epsilon \ln(1\pm \zeta)
        + (3\!-\!5\epsilon) - \f{2(2\!-\!3\epsilon)}{1\pm\zeta}
        + \f{1\!-\!\epsilon}{(1\pm\zeta)^2}
     \Big]
\nonumber\\
 \square [H^2] &=& (1\!-\!\epsilon)H^4 (6\epsilon)
\nonumber\\
 \square [H^2\ln(\zeta)] &=& (1\!-\!\epsilon)H^4
     \Big[
          6\epsilon \ln(\zeta)
        + (3\!-\!5\epsilon)
     \Big]
\nonumber\\
 \square [H^2\zeta^\omega] &=& (1\!-\!\epsilon)H^4
     \Big[
          6\epsilon
        + (3\!-\!5\epsilon)\omega
        - (1\!-\!\epsilon)\omega ^2
     \Big]
\,.
\label{box identity 3}
\end{eqnarray}
From these identities we can also obtain how the d'Alembertian acts on
the logarithms
in~(\ref{prop:full:resum:accel}--\ref{prop:full:resum:decel})
\begin{eqnarray}
 \square\bigg[H^2\ln\bigg(\f{\mu^2}{H^2}\Big(1-\f{1}{\zeta}\Big)^2\bigg)\bigg]
&=&
    (1\!-\!\epsilon)H^4
     \bigg\{
          6\epsilon \ln\Big[\f{\mu^2}{H^2}\Big(1-\f{1}{\zeta}\Big)^2\Big]
\nonumber\\
&&\hskip 2cm
        -\, \f{2\epsilon(3\!-\!5\epsilon)}{1\!-\!\epsilon}
        - \f{4(2\!-\!3\epsilon)}{1-\zeta}
        + \f{2(1\!-\!\epsilon)}{(1-\zeta)^2}
     \bigg\}
\nonumber\\
 \square \bigg[H^2 \ln\bigg(\f{\mu^2}{H^2(1+\zeta)^2}\bigg)\bigg]
&=&
    (1\!-\!\epsilon)H^4
     \bigg\{
          6\epsilon \ln\bigg[\f{\mu^2}{H^2(1+\zeta)^2}\bigg]
\nonumber\\
&&\hskip 2cm
        -\, \f{2(3\!-\!5\epsilon)}{1\!-\!\epsilon}
        + \f{4(2\!-\!3\epsilon)}{1+\zeta}
        - \f{2(1\!-\!\epsilon)}{(1+\zeta)^2}
     \bigg\}
\,.
\label{box identity 4}
\end{eqnarray}
These expressions are used in
section~\ref{The equation of state of the quantum fluid} to calculate
the trace of the stress energy tensor from the coindident propagator.

\section*{Appendix B: Integrating the stress energy conservation equation}
\label{Appendix B}

 The covariant conservation of the stress energy tensor
 for quantum fluctuations in FLRW space-times
 acquires in $D=4$ the form~(\ref{conservation}):
\begin{equation}
    \f{1}{H}\dot\rho_q + 4\rho_q= \rho_q-3p_q = -T_q
\,.
\label{conservation:AppB}
\end{equation}
When the tress energy trace $T_q$ is a function of the Hubble parameter $H$
only, we have $d/dt = -\epsilon H d/dH$, such that
equation~(\ref{conservation:AppB}) can be integrated,
\begin{equation}
  \rho_q = \f{H^{4/\epsilon}}{\epsilon}\int^H\f{d\tilde H}{H^{1+4/\epsilon}}T_q(\tilde H)
\,.
\label{AppB:2}
\end{equation}
If, on the other hand, $T_q=H^4\tau_q(\zeta)$ with $\rho_q=H^4r_q(\zeta)$,
Eq.~(\ref{conservation:AppB}) can be recast as,
\begin{equation}
    \Big(4-\zeta \f{d}{d\zeta}\Big)r_q = -\f{\tau_q}{1-\epsilon}
\,.
\label{AppB:3}
\end{equation}
This can be easily integrated to yield,
\begin{equation}
  \rho_q = H^4\f{\zeta^4}{1-\epsilon}
          \int^\zeta\f{d\tilde \zeta}{\tilde\zeta^5}\tau_q(\tilde\zeta)
\,,\qquad  p_q = \f13(\rho_q+T_q)
\,.
\label{AppB:4}
\end{equation}
The simple useful examples that make use of Eq.~(\ref{AppB:3}) are
\begin{eqnarray}
 T_q &=& t_0 H^4
   \;\Rightarrow\;
 \rho_q = t_0H^4\Big(-\f{1}{4(1-\epsilon)}\Big)
\nonumber\\
 T_q &=& t_1 H^4 \ln(H^2)
   \;\Rightarrow\;
 \rho_q = t_1H^4\Big(-\f{1}{4(1-\epsilon)}\Big)\Big(\ln(H^2)
          + \f{\epsilon}{2(1-\epsilon)}\Big)
\,.
\label{AppB:5}
\end{eqnarray}
Useful examples which make use of the integral~(\ref{AppB:4}) are,
\begin{eqnarray}
 T_q &=& t_2 H^4 \zeta^\omega
   \;\Rightarrow\;
 \rho_q = t_2 \f{H^4\zeta^\omega}{(1-\epsilon)(\omega-4)} \,,\qquad (\omega\neq 4)
\nonumber\\
 T_q &=& t_3 H^4 \zeta^4
   \;\Rightarrow\;
 \rho_q = -t_3 \f{H^4}{1-\epsilon}\zeta^4\ln(\zeta)
 \nonumber\\
 T_q &=& t_4 H^4 \ln(\zeta)
   \;\Rightarrow\;
 \rho_q = t_4H^4\Big(-\f{1}{4(1-\epsilon)}\Big)\Big(\ln(\zeta)+\f14\Big)
\,.
\label{AppB:6}
\end{eqnarray}
Here $t_n$ ($n=0,1,2,3,4$) are constants, which in general depend on
$\epsilon$ and $\xi$ but not on time.

\section*{The special cases when $\nu=3/2$ and $\nu=5/2$}
 \label{The special cases nu=3/2 and 5/2}

 The expressions~(\ref{acc:Tq:2}--\ref{acc:pq:2})
and~(\ref{dec:rhoq:2}--\ref{dec:pq:2}) are singular in the limit
when $\nu=3/2$, indicating that the $\nu=3/2$ case requires a
special attention. We shall first consider the accelerating case,
and then the decelerating case. Furthermore,
Eqs.~(\ref{dec:rhoq:3}--\ref{dec:pq:3}) are singular in the limit
when $\epsilon=3/2$. This singular behavour can be traced back to
the $\nu=5/2$ divergence in Eqs.~(\ref{rhoq2:exact:dec}) (see the
$n=1$ member of the sum~(\ref{rhoq:dec:nu-powers})). This special
case is also considered below.

 \subsubsection{The special case when $\epsilon<1$, $\nu=3/2$}
 \label{The special case epsilon<1  nu=3/2}

 In order to get the correct stress energy tensor for $\nu=3/2$
we need to take the $\nu=3/2$ limit of Eqs.~(\ref{Tq2:exact})
and~(\ref{rhoq2:exact}). This amounts to adding the corresponding
contributions from~(\ref{acc:Tq}--\ref{acc:pq})
and~(\ref{acc:Tq:2}--\ref{acc:pq:2}) and the $(\pi/2)\tan(\pi\nu)$
from~(\ref{Tq2:exact}) and~(\ref{rhoq2:exact}). The result is,
\begin{eqnarray}
  T_q &\stackrel{\zeta\rightarrow 0}{\longrightarrow}&
  -\f{(1\!-\!6\xi)^2(1\!-\!\epsilon)(2\!-\!\epsilon)H^4}{16\pi^2}
\bigg\{6\epsilon\bigg[ \ln\Big(\f{H_0\,a}{\hat H\hat a}\Big)
                  + \f{\epsilon}{4(1\!-\!\epsilon)}
                \bigg] - (3\!-\!5\epsilon)
\bigg\} \label{spec:nu32:acc:Tq}
\\
  \rho_q &\stackrel{\zeta\rightarrow 0}{\longrightarrow}&
  \f{(1\!-\!6\xi)^2(2\!-\!\epsilon)H^4}{64\pi^2}
\bigg\{6\epsilon\ln\Big(\f{H_0\,a}{\hat H\hat a}\Big)  -
(3\!-\!5\epsilon) \bigg\} \label{spec:nu32:acc:rhoq}
\\
  p_q &\stackrel{\zeta\rightarrow 0}{\longrightarrow}&
  \f{(1\!-\!6\xi)^2(2\!-\!\epsilon)H^4}{64\pi^2}
\bigg\{\Big(-1+\f43\epsilon\Big)
   \bigg[6\epsilon\ln\Big(\f{H_0\,a}{\hat H\hat a}\Big)- (3\!-\!5\epsilon)
         \bigg]
                     -2\epsilon^2
\bigg\} \,, \label{spec:nu32:acc:pq}
\end{eqnarray}
where $H_0$ is defined in~(\ref{acc:H0}). Taking a ratio of
(\ref{spec:nu32:acc:rhoq}) and~(\ref{spec:nu32:acc:pq}) gives the
equation of state parameter for this case,
\begin{eqnarray}
  w_q  &\stackrel{\zeta\rightarrow 0}{\longrightarrow}&
   -1 + \f43\epsilon - \f{2\epsilon^2}
                       {6\epsilon\ln[H_0\,a/(\hat H\hat a)]-(3\!-\!5\epsilon)}
\nonumber\\
       &\rightarrow& -1 + \f43\epsilon = w_b + \f23 \epsilon
\,. \label{spec:nu32:acc:wq}
\end{eqnarray}
Notice that this result agrees (in the limit when $a\rightarrow
\infty$) with both $w_q$ in Eq.~(\ref{acc:wq}) and with
$\nu\rightarrow 3/2$ limit -- or equivalently the $\xi\rightarrow
\epsilon(3-2\epsilon)/[6(2-\epsilon)]$ limit -- of
relation~(\ref{acc:wq:2}). Hence, when it comes to the equation of
state parameter $w_q$ at late times, there is nothing special
about the point $\nu=3/2$: the curves shown in
figures~\ref{fig1}-\ref{fig3} are continuous at $\nu=3/2$. The
only special point is the logarithmic form of the stress energy
tensor, as can be seen in
Eqs.~(\ref{spec:nu32:acc:Tq}--\ref{spec:nu32:acc:pq}). One can
check that the results identical
to~(\ref{spec:nu32:acc:Tq}--\ref{spec:nu32:acc:wq}) can be
obtained by calculating the stress energy tensor from the
$\nu=3/2$ case of the half-integer coincident
propagator~(\ref{prop:full}), representing a check of
Eqs.~(\ref{spec:nu32:acc:Tq}--\ref{spec:nu32:acc:pq}), as well as
a check of our procedure based on analytic continuation. Notice
that the logarithms drop out in the de Sitter limit when
$\epsilon\rightarrow 0$, which is a well known one loop result.
The logarithms are, however, expected to re-appear at two or
higher loop order also in de Sitter space both in massless scalar
theories~\cite{OW} as well as in quantum
gravity~\cite{Tsamis:1996qm}.

 \subsubsection{The special case when $\epsilon>1$, $\nu=3/2$}
 \label{The special case epsilon>1  nu=3/2}

Just as in the accelerating case, when matching onto deceleration
the limit $\nu\rightarrow 3/2$ appears singular, as can be seen
from Eqs.~(\ref{dec:rhoq:2}--\ref{dec:pq:2}). The full
expressions~(\ref{Tq2:exact:dec}) and~(\ref{rhoq2:exact:dec}) are
of course regular (thanks for the $\tan(\pi\nu)$ terms). The get
the correct late time limit in this case, it suffices to add the
$\tan(\pi\nu)$ terms from ~(\ref{Tq2:exact:dec})
and~(\ref{rhoq2:exact:dec}) to
Eqs.~(\ref{dec:rhoq:2}--\ref{dec:pq:2})) and take the
$\nu\rightarrow 3/2$ limit. The results are finite and -- just in
the accelerating case~\ref{The special case epsilon<1  nu=3/2} --
they acquire logarithms:
\begin{eqnarray}
  T_q &\stackrel{\zeta\rightarrow \infty}{\longrightarrow}&
  \f{(1\!-\!6\xi)^2(1\!-\!\epsilon)(2\!-\!\epsilon)^2H^4}{32\pi^2}\,\zeta^4
\qquad (\nu=3/2) \label{special:nu32:dec:Tq}
\\
  \rho_q &\stackrel{\zeta\rightarrow \infty}{\longrightarrow}&
  \f{(1\!-\!6\xi)^2(2\!-\!\epsilon)^2H^4}{32\pi^2}\, \zeta^4\ln(\zeta)
\label{special:nu32:dec:rhoq}
\\
  p_q &\stackrel{\zeta\rightarrow \infty}{\longrightarrow}&
  \f{(1\!-\!6\xi)^2(2\!-\!\epsilon)^2H^4}{32\pi^2}\, \f{\zeta^4}{3}
               \Big(\ln(\zeta)+(1\!-\!\epsilon)\Big)
\,. \label{special:nu32:dec:pq}
\end{eqnarray}
 The equation of state parameter $w_q$ follows trivially
from these relations,
\begin{eqnarray}
   w_q &\stackrel{\zeta\rightarrow \infty}{\longrightarrow}&
        \f13 - \f{\epsilon\!-\!1}{\ln(\zeta)}
         \rightarrow \f13
\,.
\label{special:nu32:dec:wq}
\end{eqnarray}
This relativistic fluid scaling is in accordance with the $\nu =
3/2$ limit of Eq.~(\ref{dec:wq:2}). Curiously $w_q$
in~(\ref{special:nu32:dec:wq}) does not depend on $\epsilon$,
implying that as long as the relation $\xi =
(3\epsilon-2)/[6(2-\epsilon)]$ holds, $w_q=1/3$ will not depend on
$\epsilon$. Just as above, it is worth commenting that results
identical to~(\ref{special:nu32:dec:Tq}) can be obtained by
calculating the stress energy tensor from the half-integer
coincident propagator~(\ref{prop:full}) by taking $\nu=3/2$ and
assuming $\epsilon>1$. In the minimally coupled scalar case when
$\xi=0$, the $\nu=3/2$ case corresponds to matter era,
$\epsilon=3/2$. In this case the logarithmic one-loop structure
exhibited in
Eqs.~(\ref{special:nu32:dec:Tq}--\ref{special:nu32:dec:pq}) is
also well known in literature.

 \subsubsection{The special case when $\nu=5/2$, $\epsilon=3/2$}
 \label{The special case nu=5/2 epsilon=3/2  }

 The last special case that requires attention is the $\nu\rightarrow 5/2$
limit on the boundary of stability, where $\xi =
-(\epsilon-1)/[3(2-\epsilon)]$ as in Eq.~(\ref{nu,xi:cr}) and
$\epsilon=3/2$, such that Eqs.~(\ref{dec:rhoq:3}--\ref{dec:pq:3})
appear singular. This limit corresponds to the $\nu=3/2$, $n=1$
term in Eq.~(\ref{rhoq:dec:nu-powers}). Similarly as above, in
this limit one gets a finite result when the term $\propto
\zeta^4\pi\tan(\pi\nu)$ from Eq.~(\ref{rhoq2:exact:dec}) is
included. In this case one reproduces the $\nu=3/2$ case discussed
above in Eqs.~(\ref{special:nu32:dec:Tq}) with
$\epsilon\rightarrow 3/2$:
\begin{eqnarray}
  T_q &\stackrel{\zeta\rightarrow \infty}{\longrightarrow}&
  -\f{(1\!-\!6\xi)^2H^4}{256\pi^2}\,\zeta^4
\qquad (\nu=5/2,\,\epsilon=3/2) \label{special:nu52:dec:Tq}
\\
  \rho_q &\stackrel{\zeta\rightarrow \infty}{\longrightarrow}&
  \f{(1\!-\!6\xi)^2H^4}{128\pi^2}\, \zeta^4\ln(\zeta)
\label{special:nu52:dec:rhoq}
\\
  p_q &\stackrel{\zeta\rightarrow \infty}{\longrightarrow}&
  \f{(1\!-\!6\xi)^2(2\!-\!\epsilon)^2H^4}{128\pi^2}\, \f{\zeta^4}{3}
               \Big(\ln(\zeta)-\f12\Big)
\,. \label{special:nu52:dec:pq}
\end{eqnarray}
 The equation of state parameter is then,
\begin{eqnarray}
   w_q &\stackrel{\zeta\rightarrow \infty}{\longrightarrow}&
        \f13 - \f{1}{2\ln(\zeta)}
         \rightarrow \f13
\,.
\label{special:nu52:dec:wq}
\end{eqnarray}
We have thus shown that a finite answer is obtained for all values
of $\nu$.

\section*{Appendix C: Explicitly calculating the $H^4 \zeta^4$ contribution}
\label{Appendix C} In this appendix we shall show that the
contribution to the stress-energy tensor proportional to
$H^4\zeta^4$ is actually ultraviolet divergent. The finite
contribution proportional to $H^4\zeta^4$ will therefore -- after
renormalization -- always be undetermined, until it is fixed by a
measurement. Thus, not only that our procedure to calculate
$\rho_q$ and $p_q$ from the trace $T_q$ in section (\ref{The
equation of state of the quantum fluid}) does not determine this
constant uniquely, but also that in the given model the sudden
matching at $t=\hat{t}$ generates an infinite amount of conformal
fluctuations and therefore it \emph{cannot} be determined
uniquely. In order to show this, we use (\ref{TMN_full}) to obtain
\begin{equation}
\langle\Omega|T_{00}|\Omega\rangle+\f{1}{D-2}\langle\Omega|T|\Omega\rangle=\Bigg(\partial_t\partial_{\tilde{t}}+\xi(R_{00}-\nabla_t\partial_t)+\f{1}{D-2}\xi\Box\Bigg)i\Delta(x;\tilde{x})\Bigg|_{x=\tilde{x}},
\end{equation}
where $T=T^\mu{}_\mu$. Using (\ref{Tmn:trace}) we then find
\begin{equation}\label{appc1}
\langle\Omega
|T_{00}|\Omega\rangle+\f{1-4\xi}{D-2-4\xi(D-1)}\langle\Omega|T|\Omega\rangle=\Bigg(\partial_t\partial_{\tilde{t}}+\xi(R_{00}-\nabla_t\partial_t)\Bigg)i\Delta(x;\tilde{x})\Bigg|_{x=\tilde{x}}.
\end{equation}
Now this equation depends also on $T_{00}$ and therefore has a
well defined contribution proportional to $H^4\zeta^4$, which we
could not determine before, when we only calculated the
contribution proportional to $T$. We shall first consider the
first term of the RHS
\begin{equation}\label{offcoinc}
\partial_t\partial_{\tilde{t}}i\Delta(x;\tilde{x})\Bigg|_{x=\tilde{x}} = \f{1}{2^{D-2}\pi^{\f{D-1}{2}}\Gamma(\f{D-1}{2})}\int dk k^{D-2}|\partial_t\psi(t,k)|^2.
\end{equation}
Now for the purpose of this section, we shall only consider the
ultraviolet divergent contribution to (\ref{offcoinc}). We saw in
section (\ref{s_prop}) that away from the matching point, only
those terms that are polynomial in $k$ contribute to the UV
divergence. Thus, using (\ref{mode_sol}), we see that we can write
\begin{equation}
|\partial_t\psi(t,k)|^2\underrightarrow{\quad\mathrm{  UV-div
}\quad} (|\alpha|^2+|\beta|^2)|\partial_t
(a(t)^{1-\f{D}{2}}u(t,k))|^2
\end{equation}
Using similar techniques that led to Eq. (\ref{int}) we find in
this case
\begin{equation}\label{coincint}
    \begin{split}
        \partial_t\partial_{\tilde{t}}i\Delta(x;\tilde{x})\Bigg|_{x=\tilde{x}\, ,\,UV}&= \f{a^{-D}}{2^{D-2}\pi^{\f{D-1}{2}}\Gamma(\f{D-1}{2})}\int dk
        k^{D-1}\\
        &\sum_{q,r,m,n=0}^{\nu-\f{1}{2}}
            \Bigg\{\f{\Gamma(\nu-\f{1}{2}+n)\Gamma(\nu+\f{1}{2}+m)\Gamma(\nu+\f{1}{2}+q)\Gamma(\nu-\f{1}{2}+r)}
            {\Gamma[\nu+\f{3}{2}-n)\Gamma(\nu+\f{1}{2}-m)\Gamma(\nu+\f{1}{2}-q)\Gamma(\nu+\f{3}{2}-r)}\\
            &\times\Bigg\{\f{1}{256}\Big(2-i(m-q)(1-\epsilon)\f{p}{k}+m
            q(1-\epsilon)^2\f{p^2}{q^2}\Big)\\
            &\times\Big(4(1-4n^2-4\nu^2)(1-4r^2-4\nu^2)\\
            &+8i(n-r)(D-1-\epsilon)\big((1-2n)(1-2r)-4(1-2n-2r)\nu^2\big)\f{aH}{k}\\
            &+
            (D-1-\epsilon)^2\big((1-2n)^2-4\nu^2\big)\big((1-2r)^2-4\nu^2\big)\f{a^2
            H^2}{k^2}\Big)\Bigg\}(-1)^{-q-n}\\
           &\Big(\f{2ik}{(1-\epsilon)p}\Big)^{-m-q}\Big(\f{2ik}{(1-\epsilon)aH}\Big)^{-n-r}\f{1}{n!m!q!r!}\Bigg\},
\end{split}
\end{equation}
where as before $p=a(\hat{t}\,) H(\hat{t}\,)$. To obtain the UV
divergent terms, we sum the first four terms of (\ref{coincint})
to obtain
\begin{equation}
    \begin{split}
    \partial_t\partial_{\tilde{t}}i\Delta(x;\tilde{x})&\Bigg|_{x=\tilde{x}\, ,\,UV}=\f{a^{-D}}{2^{D-2}\pi^{\f{D-1}{2}}\Gamma(\f{D-1}{2})}\int
    dk
    k^{D-1}\\
    &\Bigg\{\f{1}{2}+\f{1}{16}\Big(9-2\epsilon+\epsilon^2+2D(D-4)-4(1-\epsilon)^2\nu^2\Big)\f{a^2H^2}{k^2}\\
    &+\f{1}{64}(1-\epsilon)^2(\nu^2-\f{1}{4})\Big[4(1-\epsilon)^2\Big(\nu^2-\f{1}{4}\Big)\zeta^4\\
    &+\big(73+4D^2-16D(2-\epsilon)-82\epsilon+25\epsilon^2-4(1-\epsilon)^2\nu^2\big)\Big]\f{a^4H^4}{k^4}\Bigg\},
\end{split}
\end{equation}
where $\zeta=\f{p}{aH}$. The powerlaw divergences ($k^{D-1}$ and
$k^{D-3}$) are automatically subtracted in dimensional
regularization and we are thus left with the logarithmic
divergence. We evaluate this divergence as in section \ref{sec_UV}
to obtain for the $1/(D-4)$ part
\begin{equation}
    \f{(2-\epsilon)(1-6\xi)}{32\pi^2(D-4)}\Bigg(3(\epsilon(1-2\epsilon)-2\xi(2-\epsilon))-(2-\epsilon)(1-6\xi)\zeta^4\Bigg)\mu^{D-4}H^4,
\end{equation}
where we used the expression for $\nu$ from (\ref{nu}). The
$1/(D-4)$ part from the other two terms from the RHS of
(\ref{appc1}) are easily evaluated, using (\ref{propUV}) and
\begin{equation}
    \begin{split}
    R_{00} &= -3(1-\epsilon)H^2 \\
    \nabla_t\partial_t H^2 &= 6\epsilon^2 H^4.
    \end{split}
\end{equation}
We can put all terms together and obtain for the $1/(D-4)$
contribution to the RHS of (\ref{appc1})
\begin{equation}\label{new}
\f{(2-\epsilon)(1-6\xi)}{32\pi^2(D-4)}\Bigg(3\epsilon\Big(1-2\epsilon-2\xi(1-4\epsilon)\Big)-(2-\epsilon)(1-6\xi)\zeta^4\Bigg)H^4.
\end{equation}
Thus we indeed find that the contribution comes in two parts: one
proportional to $H^4$, and one proportional to $H^4\zeta^4$. As a
check, we can calculate the LHS of (\ref{appc1}), using the
ultraviolet contributions we obtained for the trace in
(\ref{Tq:div}). using the conservation equation we can then find,
using (\ref{AppB:2}) also the divergent contribution to the energy
density, since if $\langle\Omega|T|\Omega\rangle\propto H^4$ we
have, apart from a possible $H^4\zeta^4$ contribution, that
\begin{equation}
    \langle\Omega|T_{00}|\Omega\rangle=-\f{1}{4(1-\epsilon)}\langle\Omega|T|\Omega\rangle .
\end{equation}
Using (\ref{Tq:div}) we then thus find for the $1/(D-4)$ part of
the LHS of \ref{appc1}
\begin{equation}
\f{(2-\epsilon)(1-6\xi)}{32\pi^2(D-4)}\Bigg(3\epsilon\Big(1-2\epsilon-2\xi(1-4\epsilon)\Big)\Bigg)H^4.
\end{equation}
In other words, the calculation leading to (\ref{new}) is
consistent with the calculation in the rest of this paper, apart
from the $H^4\zeta^4$ term. From (\ref{new}) we thus see that
there is a divergence $\propto H^4\zeta^4/(D-4)$ contributing to
the stress-energy tensor. This divergence could be subtracted by a
counterterm of the form
\begin{equation}\label{rad_counterterm}
    \alpha\int d^Dx\sqrt{-g}p_r,
\end{equation}
where $\alpha$ is a constant and $p_r$ is the pressure of some
radiation fluid, obeying $p_r=\f{1}{3} \rho_r$. This fluid could
for example be a photon fluid, or a scalar field fluid in thermal
equilibrium. This renormalization is not completely satisfactory,
since it requires the addition of a new field, not present in the
original model. However, one needs to keep in mind here that our
model is incomplete anyway. The sudden matching is put in by hand,
where in a realistic model, it should arise from the dynamics of
fields. Moreover, in a realistic model, the matching is never
instantaneous, which should remove this UV divergence anyway.
Therefore we do not feel that these pathologies, arising from the
sudden matching are problematic. However, given the fact that the
present model needs a counterterm like (\ref{rad_counterterm}),
the undetermined finite part contributing to the counterterm makes
any $H^4\zeta^4$ contribution to the stress energy tensor
arbitrary.

\end{document}